\documentclass[english,amssymb,amsmath,showpacs,twocolumn,superscriptaddress]{revtex4-1} 
\usepackage[english]{babel}
\usepackage[utf8]{inputenc}
\usepackage{amssymb,amsmath}
\usepackage[colorlinks,pdfpagelabels,pdfstartview = FitH,bookmarksopen = true,bookmarksnumbered = true,linkcolor = black,plainpages = false,hypertexnames = false,citecolor = black]{hyperref}
\usepackage{mathtools}
\usepackage{textcomp}
\usepackage{overpic}

\usepackage{ulem}

\begin{document}
	\title{Theory of the Coherent Response of Magneto-Excitons and Magneto-Biexcitons \\ in Monolayer Transition Metal Dichalcogenides}
	
	\author{Florian Katsch}
	\affiliation{Institut f\"ur Theoretische Physik, Nichtlineare Optik und Quantenelektronik, Technische Universit\"at Berlin, 10623 Berlin, Germany}
	\author{Dominik Christiansen}
	\affiliation{Institut f\"ur Theoretische Physik, Nichtlineare Optik und Quantenelektronik, Technische Universit\"at Berlin, 10623 Berlin, Germany}
	\author{Robert Schmidt}
	\affiliation{Institute of Physics and Center for Nanotechnology, University of Münster, 48149 Münster, Germany}
	\author{Steffen Michaelis de Vasconcellos}
	\affiliation{Institute of Physics and Center for Nanotechnology, University of Münster, 48149 Münster, Germany}
	\author{Rudolf Bratschitsch}
	\affiliation{Institute of Physics and Center for Nanotechnology, University of Münster, 48149 Münster, Germany}
	\author{Andreas Knorr}
	\affiliation{Institut f\"ur Theoretische Physik, Nichtlineare Optik und Quantenelektronik, Technische Universit\"at Berlin, 10623 Berlin, Germany}
	\author{Malte Selig}
	\affiliation{Institut f\"ur Theoretische Physik, Nichtlineare Optik und Quantenelektronik, Technische Universit\"at Berlin, 10623 Berlin, Germany}
	
	\begin{abstract}
		The recent accessibility of high quality, charge neutral monolayer transition metal dichalcogenides with narrow exciton linewidths at the homogeneous limit provides an ideal platform to study excitonic many-body interactions. In particular, the possibility to manipulate coherent exciton-exciton interactions, which govern the ultrafast nonlinear optical response, by applying an external magnetic field has not been considered so far. We address this discrepancy by presenting a nonlinear microscopic theory in the coherent limit for optical excitations in the presence of out-of-plane, in-plane, and tilted magnetic fields. Specifically, we explore the magnetic-field-induced exciton and biexciton fine structure and calculate their oscillator strengths based on a Heisenberg equations of motion formalism. Our microscopic evaluations of pump-probe spectra allow to interpret and predict coherent signatures in future wave-mixing experiments.
	\end{abstract}
	
	\maketitle
	
\section{Introduction}

		Monolayer transition metal dichalcalcogenides (TMDCs) exhibit outstanding electronic and optical properties \cite{splendiani2010emerging,mak2010atomically} including excitons (bound electron-hole pairs) with exceptionally large binding energies \cite{mak2013tightly,berkelbach2013theory,chernikov2014exciton,he2014tightly}.
		The TMDC band structure is characterized by direct band gaps with strong spin-orbit interaction leading to a spin-splitting of valence and conduction bands at the non-equivalent corners K and K$'$ of the first Brillouin zone \cite{kuc2011influence,zhu2011giant,wang2012electronics,kumar2012electronic,yun2012thickness,druppel2018electronic}.
		The spin-splitting together with the valley selective circular dichroism of monolayer TMDCs allows to separately access the valley and spin degree of freedom.
		The complex band structure introduces a variety of distinct exciton configurations \cite{deilmann2019finite,yu2019exploration,xie2019theory} as well as related trion \cite{druppel2017diversity,deilmann2017dark,florian2018dielectric,tempelaar2019many,arora2019excited,arora2019dark} and biexciton configurations \cite{steinhoff2018biexciton,kuhn2019combined,katsch2020theory}.
		Gates, barriers, or common accidental impurities, resulting in doped TMDC samples with pronounced trions besides neutral excitons, motivated numerous experimental investigations of the exciton and trion dynamics \cite{zhu2014exciton,kumar2014valley,mai2014exciton,yan2015valley,wang2015polarization,dal2015ultrafast,singh2016trion,plechinger2016trion,schmidt2016ultrafast,smolenski2016tuning,plechinger2017valley,mccormick2017imaging,tang2019single}.
		In contrast, experimental investigations of biexcitons are more involved because biexciton resonances are hard to resolve in spectroscopic experiments performed on monolayer TMDCs \cite{hao2017neutral,steinhoff2018biexciton} due to the small biexciton binding energy \cite{zhang2015excited,mayers2015binding,kylanpaa2015binding,kidd2016binding,mostaani2017diffusion,kezerashvili2017trion,szyniszewski2017binding,van2018excitons,kuhn2019tensor} compared to the large exciton linewidth \cite{moody2016exciton,selig2016excitonic,christiansen2017phonon,lengers2020theory}.
		However, recent advance in encapsulating monolayer TMDCs in hexagonal boron nitride (hBN) demonstrated to dramatically decrease the exciton linewidth down to the homogeneous limit
		resulting in spectrally sharp exciton resonances \cite{cadiz2017excitonic,ajayi2017approaching,wierzbowski2017direct,robert2018optical,martin2018encapsulation,fang2019control,jakubczyk2019coherence}.
		Therefore, the encapsulation of TMDCs in hBN together with an externally applied gate voltage can effectively suppresse highly charge tunable features like trions and allows to accomplish intrinsic TMDC samples approaching the homogeneous limit at cryogenic temperatures where charge neutral biexcitons are significant \cite{chen2018coulomb,ye2018efficient,barbone2018charge,li2018revealing,li2019momentum,paur2019electroluminescence}.
		Whereas the energetically highest valence and lowest conduction bands near the K and K$'$ points are symmetric except for opposite spins, this symmetry is broken by external magnetic fields:
		Out-of-plane magnetic fields (oriented perpendicular to the monolayer plane in a Faraday geometry) introduce different valley and spin-dependent Zeeman shifts of the exciton energies at the K and K$'$ points \cite{li2014valley,srivastava2015valley,macneill2015breaking,wang2015magneto,aivazian2015magnetic,stier2016exciton,stier2016probing,plechinger2016excitonic,mitioglu2016magnetoexcitons,arora2016valley,scrace2015magnetoluminescence,schmidt2016magnetic,wang2016control,nagler2018zeeman,zipfel2018spatial,wang2018strongly,koperski2018orbital,arora2018zeeman,goryca2019revealing,zhang2019zeeman,li2019emerging,deilmann2020ab,wozniak2020exciton}, cf. Fig.~\ref{Bild-AB}\,(a).
		On the other hand, in-plane magnetic fields (oriented parallel to the monolayer plane in a Voigt geometry) soften the optical selection rules and lead to a brightening of spin-forbidden excitons with increasing magnetic fields \cite{zhang2017magnetic,molas2017brightening,van2018strong,lu2019magnetic,robert2020measurement,feierabend2020brightening}.
		Hence, exposing TMDCs to an external magnetic field is expected to represent an ideal platform to study Coulomb many-body interactions in coherent pump-probe experiments performed on high quality monolayer TMDCs at cryogenic temperatures.
		Here, exciton-exciton scattering and a rich biexciton fine structure are expected to govern the ultrafast nonlinear optical response.
		Whereas previous experimental studies concentrated on incoherent photoluminescence measurements \cite{chen2018coulomb,ye2018efficient,barbone2018charge,li2018revealing,li2019momentum,paur2019electroluminescence}, we propose to gain a new perspective of exciton-exciton interaction and the biexciton dynamics via our theoretical analysis in ultrafast pump-probe spectroscopy.
		We demonstrate that the pump-probe spectra mirror the excitonic Zeeman shifts in the presence of an out-of-plane magnetic field.
		In particular, biexciton resonances inherit the $g$-factor of the probed exciton resonances.
		Moreover, we show pronounced nonlinear renormalizations of previously spin-forbidden dark excitons and a rich biexciton fine structre induced by an in-plane magnetic field.
		The combined influence of out-of-plane and in-plane magnetic field contributions for a tilted magnetic field allows to enhance or suppress the pump-probe response of dark excitons as well as corresponding biexciton resonances.
		This paper is organized as follows:
		We first introduce the observables including bright and dark excitons as well as biexcitons and exciton-exciton scattering continua required for a nonlinear coherent description in section~\ref{sec:observables}.
		Subsequently, in section~\ref{app:eqs-of-motion} we develop a microscopic theory based on excitonic Heisenberg equations of motion for the coherent response of monolayer TMDCs in the presence of an externally applied magnetic field.
		In section~\ref{sec:results} we separately study out-of-plane, in-plane, and tilted magnetic fields.
		At first, we summarize the magnetic-field-dependent shifted exciton resonance energies and linear transmission.
		Afterwards, we focus on the rich magnetic-field-induced biexciton landscape and the magnetic-field-induced resonance energy shifts.
		On the basis of numerical evaluations of coherent pump-probe spectroscopy, we investigate the pump-dependent changes of the exciton resonances due to exciton-exciton scattering.
		Moreover, we identify biexcitons with sufficient oscillator strength to appear in the nonlinear optical response and how the oscillator strength can be manipulated in the presence of differently orientated magnetic fields.
		Our analysis shows that coherent spectroscopy performed on hBN encapsulated TMDCs at low temperatures significantly enhances the understanding and interpretation of many-body states in monolayer TMDCs.
		Finally, we conclude in section~\ref{sec:conclusion}.

\section{Observables} \label{sec:observables}

		\begin{figure}
			\centering
			\begin{overpic}[width=0.8\columnwidth]{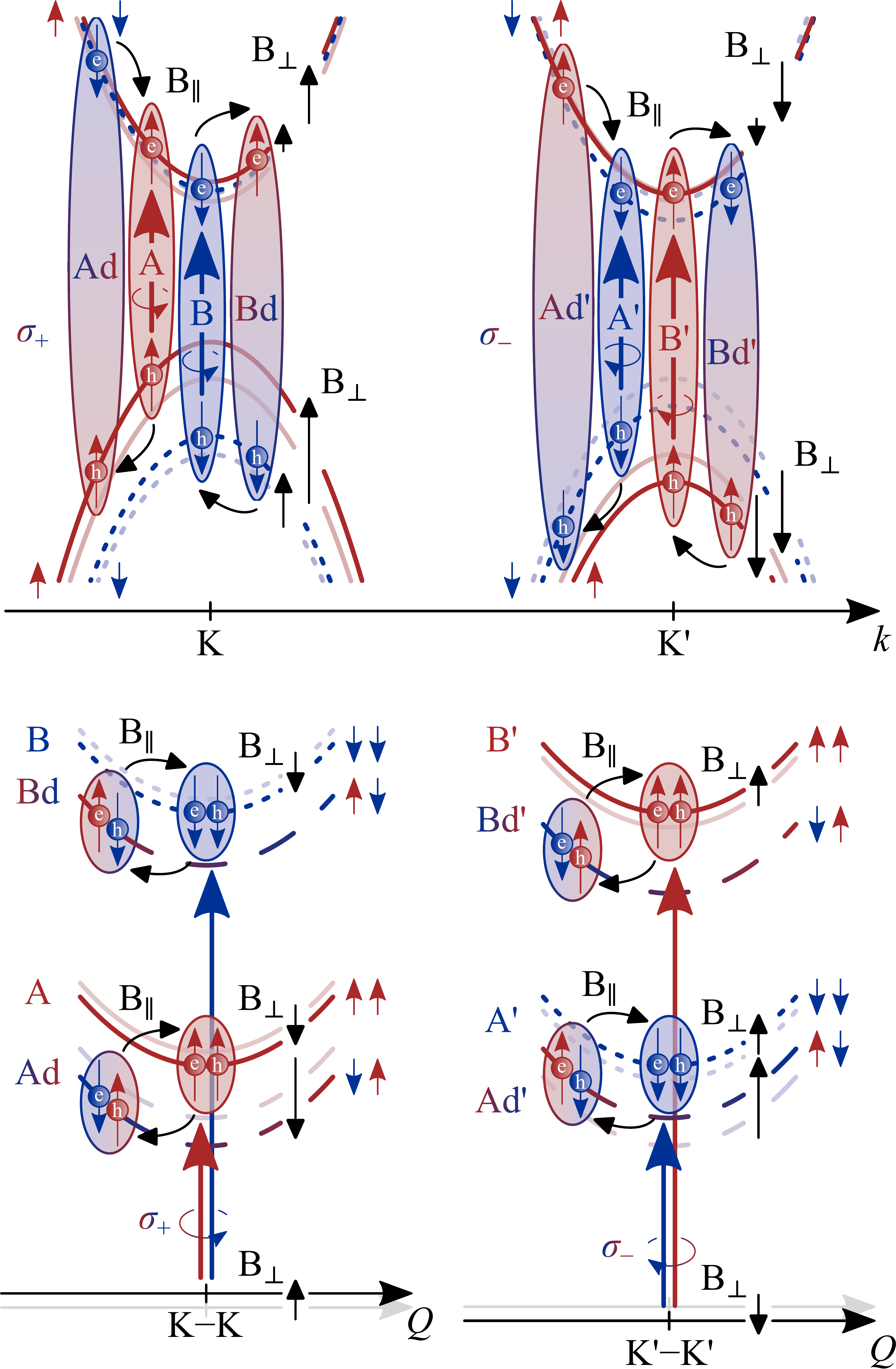}
				\put(-5,98){(a)}
				\put(-4,46){(b)}
			\end{overpic}
			\caption{\textbf{Electron-hole pairs and excitons in external magnetic fields.}
				Illustrated are (a)~electron-hole pairs in the MoS$_2$ band structure near the K and K$'$ points and (b)~exciton states associated with the K~point (K$-$K) and K$'$ point (K$'-$K$'$).
					A$^{(\prime)}$ and B$^{(\prime)}$~excitons with same electron and hole spin are optically generated by $\sigma_{+(-)}$~circularly polarized light.
					Out-of-plane magnetic fields $B_\perp$ shift the conduction and valence bands due to the valley orbital, atomic orbital, and spin magnetic moments.
					The magnetic moments shift the K$-$K and K$'-$K$'$~excitonic ground states in opposite directions (shifts of the x-axes in (b)) and shift the various bright and dark exciton states with different magnitude depending on the magnetic field $B_\perp$ in opposite directions for K$-$K and K$'-$K$'$~excitons.
					In-plane magnetic fields $B_\parallel$ couple bright A$^{(\prime)}$ and B$^{(\prime)}$~excitons to dark Ad$^{(\prime)}$ and Bd$^{(\prime)}$~excitons with opposite electron and hole spin.
			}
			\label{Bild-AB}
		\end{figure}
\begin{figure*}
			\centering
			\includegraphics[width=2.\columnwidth]{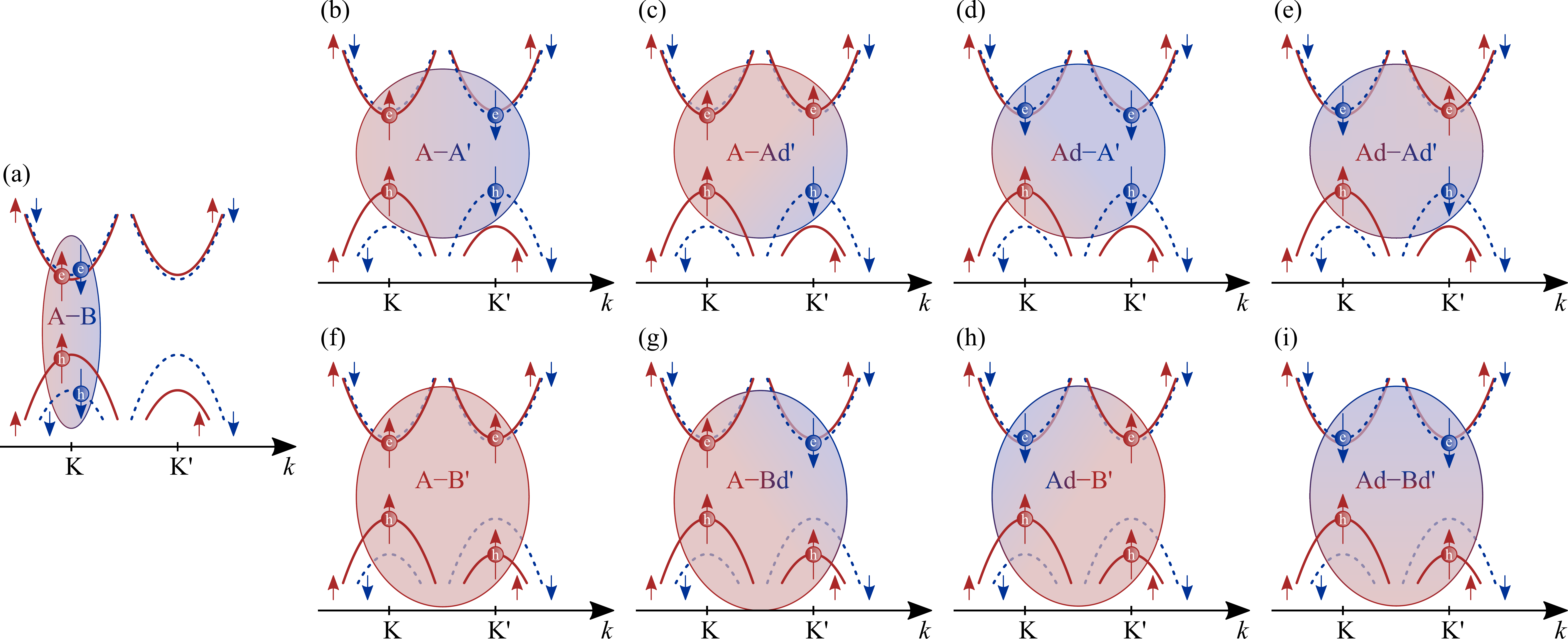}
			\caption{\textbf{Illustration of two-electron and two-hole Coulomb correlations.}
				Shown are exemplary (a)~intra- and (b$-$i)~intervalley two-electron and two-hole Coulomb correlations which exhibit bound states (biexcitons) which are later relevant for pump photon energies resonant to the A$_\text{1s}$~exciton.
				(a)~Intravalley correlations comprise two electrons and two holes near the K or K$'$ point, whereas (b$-$i)~intervalley correlations involve an electron and a hole near the K~point as well as a second electron and hole in the K$'$~valley.
			}
			\label{fig:bix-schema}
		\end{figure*}

		The optical response of monolayer TMDCs is determined by the polarization density $P\hspace{0.5mm}^{\sigma_{+(-)}}(t)$:
		\begin{align}
		P\hspace{0.5mm}^{\sigma_{\pm}}(t) = \ &\left(\delta_{+,\pm} \ \delta_{\xi,\text{K}} + \delta_{-,\pm}  \ \delta_{\xi,\text{K}'} \right) \notag \\
		& \times \sum_{s,\nu,\textbf{\textit{q}}} \varphi^*\hspace{0.5mm}^{\xi,s,s}_{\nu,\textbf{\textit{q}}} \ d\hspace{0.5mm}^{c,v}_{\xi,s} \ P\hspace{0.5mm}^{\xi,s,s}_{\nu} \notag \\
		& + c.c. \ .
		\end{align}
		Due to the valley selective circular dichroism \cite{xiao2012coupled,cao2012valley}, the polarization density $P\hspace{0.5mm}^{\sigma_{+}}(t)$ is associated with the $\xi=$~K valley and $P\hspace{0.5mm}^{\sigma_{-}}(t)$ with the $\xi=$~K$'$ valley.
		$\varphi^*\hspace{0.5mm}^{\xi,s,s}_{\nu,\textbf{\textit{q}}}$ is the exciton wave function obtained from solving the Wannier equation \cite{kira2006many} for excitons at the $\xi =$~K, K$'$ point with same electron and hole spin $s =  \ \uparrow$, $\downarrow $ and exciton quantum number $\nu = \text{1s}$, 2p, 2s, and so on.
		The dipole matrix element $d\hspace{0.5mm}^{c,v}_{\xi,s}$ is defined in Eq.~(\ref{eq:dipole-mat}).
		To work in a convenient basis set, interband transitions $\big\langle{c^\dag_{\xi,s_1,\textbf{\textit{k}}} v^{\phantom{\dag}}_{\xi,s_2,\textbf{\textit{k}}}}\big\rangle$ were expanded in terms of exciton transitions~$P\hspace{0.5mm}^{\xi,s_1,s_2}_{\nu}$:
		\begin{align}
			\big\langle{c^\dag_{\xi,s_1,\textbf{\textit{k}}} v^{\phantom{\dag}}_{\xi,s_2,\textbf{\textit{k}}}}\big\rangle = \ &  \sum_{\nu} \varphi^*\hspace{0.5mm}^{\xi,s_1,s_2}_{\nu,\textbf{\textit{k}}} \ P\hspace{0.5mm}^{\xi,s_1,s_2}_{\nu}. \label{eq:exciton-transition}
		\end{align}
		The energetically lowest exciton series associated with the $s_1 = s_2 = \ \uparrow(\downarrow)$ conduction and valence bands at the $\xi =$ K$^{(\prime)}$~point are referred to as A$^{(\prime)}$~excitons $P\hspace{0.5mm}^{\text{K},\uparrow,\uparrow}_{\nu}$ ($P\hspace{0.5mm}^{\text{K}',\downarrow,\downarrow}_{\nu}$), cf. Fig.~\ref{Bild-AB}.
		The spin-split energetically higher B$^{(\prime)}$~transitions $P\hspace{0.5mm}^{\text{K},\downarrow,\downarrow}_{\nu}$ ($P\hspace{0.5mm}^{\text{K}',\uparrow,\uparrow}_{\nu}$) are associated with the $s_1 = s_2 = \ \downarrow(\uparrow)$ conduction and valence bands at the $\xi =$ K$^{(\prime)}$~point.
		These A$^{(\prime)}$ and B$^{(\prime)}$~exciton states exhibit an in-plane dipole and are referred to as \textit{bright} excitons since they can be optically excited with $\sigma_{+(-)}$~circularly polarized light perpendicular to the monolayer plane \cite{wang2017plane,zhou2017probing,shree2020guide}.
		In contrast, \textit{dark} excitons with out-of-plane dipole are characterized by opposite electron and hole spins $s_1 \neq s_2$ and can not be optically excited with light perpendicular to the monolayer plane \cite{wang2017plane,zhou2017probing,shree2020guide}.
		The energetically lower dark transitions are referred to as Ad$^{(\prime)}$~exciton $P\hspace{0.5mm}^{\text{K},\downarrow,\uparrow}_{\nu}$ ($P\hspace{0.5mm}^{\text{K}',\uparrow,\downarrow}_{\nu}$), whereas the higher states are called Bd$^{(\prime)}$~excitons $P\hspace{0.5mm}^{\text{K},\uparrow,\downarrow}_{\nu}$ ($P\hspace{0.5mm}^{\text{K}',\downarrow,\uparrow}_{\nu}$), cf. Fig.~\ref{Bild-AB}.
		The dynamics of \textit{bright} and \textit{dark} exciton transitions~$P\hspace{0.5mm}^{\xi,s_1,s_2}_{\nu}$ is based on Heisenberg equations of motion truncated to the third order in the exciting electromagnetic field \cite{axt1994dynamics,axt1994role,lindberg1994chi,bartels1997coherent,koch2001correlation}.
		In this coherent limit valid on ultrashort timescales \cite{selig2017dark}, the exciton transitions couple to two-electron and two-hole Coulomb correlations $\big\langle{c^\dag_{\xi_1,s_1,\textbf{\textit{k}}_1+\textbf{\textit{P}}} v^{\phantom{\dag}}_{\xi_1,s_2,\textbf{\textit{k}}_1} c^\dag_{\xi_2,s_3,\textbf{\textit{k}}_2-\textbf{\textit{P}}} v^{\phantom{\dag}}_{\xi_2,s_4,\textbf{\textit{k}}_2}}\big\rangle^{\text{c}}$ expressed in the convenient basis of transitions $B\hspace{0.5mm}^{\xi_1,s_1,s_2,\xi_2,s_3,s_4}_{\pm,\mu}$ \cite{schafer2013semiconductor,takayama2002t}:
		\begin{align}
		& \big\langle{c^\dag_{\xi_1,s_1,\textbf{\textit{k}}_1+\textbf{\textit{P}}} v^{\phantom{\dag}}_{\xi_1,s_2,\textbf{\textit{k}}_1} c^\dag_{\xi_2,s_3,\textbf{\textit{k}}_2-\textbf{\textit{P}}} v^{\phantom{\dag}}_{\xi_2,s_4,\textbf{\textit{k}}_2}}\big\rangle^{\text{c}} \notag \\
		& = \sum_{\pm,\nu_1,\nu_2,\mu} \Big( \varphi^*\hspace{0.5mm}^{\xi_1,s_2}_{\nu_1,\textbf{\textit{k}}_1 +\beta_{\xi_1,s_2}\textbf{\textit{P}}} \ \varphi^*\hspace{0.5mm}^{\xi_2,s_4}_{\nu_2,\textbf{\textit{k}}_2 -\beta_{\xi_2,s_4}\textbf{\textit{P}}} \notag \\
		& \hspace{18.2mm} \times \Phi^{\text{R}}\hspace{0.5mm}^{\xi_1,s_1,s_2,\xi_2,s_3,s_4}_{\pm,\mu,\nu_1,\nu_2,\textbf{\textit{P}}} \ B\hspace{0.5mm}^{\xi_1,s_1,s_2,\xi_2,s_3,s_4}_{\pm,\mu} \notag \\
		& \hspace{18.2mm} \mp\varphi^*\hspace{0.5mm}^{\xi_1,s_2}_{\nu_1,\alpha_{\xi_1,s_2}\textbf{\textit{k}}_1 +\beta_{\xi_1,s_2}\left(\textbf{\textit{k}}_2-\textbf{\textit{P}}\right)} \notag \\
		& \hspace{21.4mm} \times \varphi^*\hspace{0.5mm}^{\xi_2,s_4}_{\nu_2,\beta_{\xi_2,s_4}\left(\textbf{\textit{k}}_1+\textbf{\textit{P}}\right) +\alpha_{\xi_2,s_4}\textbf{\textit{k}}_2} \notag \\
		& \hspace{21.4mm} \times \Phi^{\text{R}}\hspace{0.5mm}^{\xi_1,s_1,s_2,\xi_2,s_3,s_4}_{\pm,\mu,\nu_1,\nu_2,-\textbf{\textit{k}}_1+\textbf{\textit{k}}_2-\textbf{\textit{P}}} \notag \\
		& \hspace{21.4mm} \times B\hspace{0.5mm}^{\xi_1,s_1,s_2,\xi_2,s_3,s_4}_{\pm,\mu} \Big) . \label{eq:four-particle-corr}
		\end{align}
		Solving the two-electron and two-hole Wannier equation \cite{schafer2013semiconductor,takayama2002t} provides the wave functions $\Phi^{\text{R}}\hspace{0.5mm}^{\xi_1,s_1,s_2,\xi_2,s_3,s_4}_{\pm,\mu,\nu_1,\nu_2,\textbf{\textit{P}}}$ where the quantum number $\mu$ includes biexcitons (bound solutions $\mu = b$) and continuous states of the exciton-exciton scattering continuum (continuum of unbound solutions $\mu \neq b$) \cite{axt1998exciton,schumacher2005coherent,schumacher2006coherent}.
		The index $\pm$ states whether the two-electron and two-hole correlation in Eq.~(\ref{eq:four-particle-corr}) is symmetric ($+$) or anti-symmetric ($-$) under electron exchange \cite{schafer2013semiconductor}.
		In particular, only the anti-symmetric ($-$) channel exhibits bound solutions indicated by $\mu = b$ \cite{takayama2002t}.
		In case of a vanishing in-plane magnetic field, as investigated in Ref.~\cite{katsch2020theory}, only two-electron and two-hole Coulomb correlations $B\hspace{0.5mm}^{\xi_1,s_1,s_2,\xi_2,s_3,s_4}_{\pm,\mu}$ with pairwise identical electron and hole spins $s_1=s_2$ and $s_3=s_4$ need to be considered.
		However, nonzero in-plane magnetic fields break this symmetry and require to consider additional two-electron and two-hole transitions $B\hspace{0.5mm}^{\xi_1,s_1,s_2,\xi_2,s_3,s_4}_{\pm,\mu}$ with $s_1 \neq s_2$ or $s_2 \neq s_3$.
		In the following, the correlations $B\hspace{0.5mm}^{\xi_1,s_1,s_2,\xi_2,s_3,s_4}_{\pm,\mu}$ are referred to as $\{\xi_1,s_1,s_2\}$$-$$\{\xi_2,s_3,s_4\}$~Coulomb correlations.
		The denotation hints at the valleys and spins of the involved electrons and holes:
		For instance, $B\hspace{0.5mm}^{\text{K},\uparrow,\uparrow,\text{K},\downarrow,\downarrow}_{\pm,\mu}$ corresponds to the A$-$B~correlation (which is identical to the Ad$-$Bd~correlations) and $B\hspace{0.5mm}^{\text{K},\downarrow,\uparrow,\text{K}',\downarrow,\downarrow}_{\pm,\mu}$ is called Ad$-$A$'$~correlation.
		Fig.~\ref{fig:bix-schema} illustrates a selection of (a)~intravalley and (b$-$i)~intervalley $\{\xi_1,s_1,s_2\}$$-$$\{\xi_2,s_3,s_4\}$~Coulomb correlations.
		Note that the repulsive interaction of two electrons or holes in the same conduction or valence bands precludes the formation of intravalley A$-$A, B$-$B, Ad$-$B, and A$-$Bd biexcitons.

\section{Excitonic Equations of Motion} \label{app:eqs-of-motion}

		The Heisenberg equation of motion for the exciton transition $P\hspace{0.5mm}^{\xi_1,s_1,s_2}_{\nu_1}$ is given by:
		\begin{align}
		& \left[\partial_t + \gamma\hspace{0.5mm}^{\xi_1,s_1,s_2}_{\text{x}} -\frac{\text{i}}{\hbar} \left(\epsilon\hspace{0.5mm}^{\xi_1,s_1,s_2}_{\text{x},\nu_1} + \varepsilon\hspace{0.5mm}^{\xi_1,s_1,s_2}_{B_\perp} \right) \right] P\hspace{0.5mm}^{\xi_1,s_1,s_2}_{\nu_1} \notag \\
		& = - \text{i} \delta_{s_1,s_2} \Omega\hspace{0.5mm}^{\xi_1,s_1}_{\nu_1} + \partial_t \ P\hspace{0.5mm}^{\xi_1,s_1,s_2}_{\nu_1} \big|_{B_\parallel} \notag \\
		& \hspace{4mm} + \partial_t \ P\hspace{0.5mm}^{\xi_1,s_1,s_2}_{\nu_1} \big|_\text{PB}  + \partial_t \ P\hspace{0.5mm}^{\xi_1,s_1,s_2}_{\nu_1} \big|_\text{HF} \notag \\
		& \hspace{4mm} + \partial_t \ P\hspace{0.5mm}^{\xi_1,s_1,s_2}_{\nu_1} \big|_{\text{corr.}} . \label{eq:exciton-dyn}
		\end{align}
		The left-hand side of Eq.~(\ref{eq:exciton-dyn}) describes excitonic oscillations with the exciton resonance energy at zero magnetic field $\epsilon\hspace{0.5mm}^{\xi_1,s_1,s_2}_{\text{x},\nu_1}$.
		An out-of-plane magnetic field $B_\perp$ renormalizes the exciton energy by $\varepsilon\hspace{0.5mm}^{\xi_1,s_1,s_2}_{B_\perp}$, cf. Fig.~\ref{Bild-AB}.
		The Zeeman shift $\varepsilon\hspace{0.5mm}^{\xi_1,s_1,s_2}_{B_\perp} = \varepsilon\hspace{0.5mm}^{\xi_1,s_1}_{c,B_\perp} - \varepsilon\hspace{0.5mm}^{\xi_1,s_2}_{v,B_\perp}$ breaks the symmetry between the K and K$'$ points due to conduction $\varepsilon\hspace{0.5mm}^{\xi_1,s_1}_{c,B_\perp}$ and valence band shifts $\varepsilon\hspace{0.5mm}^{\xi_2,s_2}_{v,B_\perp}$ with different signs and magnitude depending on the valley~$\xi_1$ and spins~$s_1$ and $s_2$:
		\begin{align}
		\varepsilon\hspace{0.5mm}^{\xi,s}_{c,B_\perp} = \ & \Big[\left( \delta_{\xi,\text{K}} - \delta_{\xi,\text{K}'}\right) \frac{m_0}{\bar{m}}  \notag \\
		& \hspace{1.8mm}+ \left(\delta_{s,\uparrow} - \delta_{s,\downarrow}\right)\Big] \mu_B B_\perp , \label{eq:Zeeman-1} \\
		\varepsilon\hspace{0.5mm}^{\xi,s}_{v,B_\perp} = \ & \Big[ \left( \delta_{\xi,\text{K}} - \delta_{\xi,\text{K}'}\right) \left(2+\frac{m_0}{\bar{m}}\right) \notag \\
		& \hspace{1.8mm} + \left(\delta_{s,\uparrow} - \delta_{s,\downarrow}\right)\Big] \mu_B B_\perp \label{eq:Zeeman-2} .
		\end{align}
		Here, $\mu_B$ is the Bohr magneton, $m_0$ is the free electron mass, and $\bar{m} = \frac{1}{8} \sum_{\xi,s} (m^e_{\xi,s}+m^h_{\xi,s})$ is the mean effective mass of the eight band model \cite{yao2008valley,xu2014spin}.
		The latter involves the effective mass $m^{e(h)}_{\xi,s}$ of the $\{\xi,s\}$ conduction (valence) band.
		A derivation based on the underlying different magnetic moments contributing in the presence of a magnetic field (atomic orbital, valley orbital, and spin magnetic moments) is given in Appendix~\ref{app:magnetic-field}.
		Exciton-phonon interactions damp the excitonic oscillations described by Eq.~(\ref{eq:exciton-dyn}) with the microscopically calculated dephasing constant $\gamma\hspace{0.5mm}^{\xi_1,s_1,s_2}_{\text{x}}$ \cite{selig2016excitonic,khatibi2018impact}.
		On the other hand, the radiative dephasing $\gamma\hspace{0.5mm}^{\xi_1,s_1,s_2}_{\text{r}}$, which dominates the exciton dephasing of hBN encapsulated high-quality monolayer TMDCs at cryogenic temperatures \cite{cadiz2017excitonic,ajayi2017approaching,robert2018optical,martin2018encapsulation,fang2019control}, does not appear explicitly in Eq.~(\ref{eq:exciton-dyn}).
		Instead, the radiative dephasing $\gamma\hspace{0.5mm}^{\xi_1,s_1,s_2}_{\text{r}}$ directly follows from the simultaneous solution of Maxwell's wave equation together with the excitonic Bloch equations \cite{knorr1996theory,stroucken1996coherent}.
		The first contribution to the right-hand side of Eq.~(\ref{eq:exciton-dyn}) describes the optical source term for bright excitons $P\hspace{0.5mm}^{\xi_1,s_1,s_2}_{\nu_1}$ with equal electron and hole spin $s_1 = s_2$.
		The excitonic Rabi frequency $\Omega\hspace{0.5mm}^{\xi_1,s_1}_{\nu}$ depends on the dipole matrix element $d\hspace{0.5mm}^{c,v}_{\xi,s}$ and the envelope of the light field at the monolayer position $\tilde{E}^{\sigma_\pm}(t)$:
		\begin{align}
		\Omega\hspace{0.5mm}^{\xi_1,s_1}_{\nu} = \ & \left(\delta_{+,\pm} \ \delta_{\xi_1,K} + \delta_{-,\pm} \ \delta_{\xi_1,K'}\right) \notag \\
		& \times \frac{1}{\hbar}\sum_{\textbf{\textit{q}}} \varphi\hspace{0.5mm}^{\xi_1,s_1,s_1}_{\nu,\textbf{\textit{q}}} \ \big(d\hspace{0.5mm}^{c,v}_{\xi,s}\big)^* \big[\tilde{E}^{\sigma_\pm}(t)\big]^* e^{\text{i} \omega_0 t} .
		\end{align}
		The valley selection rules \cite{cao2012valley} are represented by $\left(\delta_{+,\pm} \ \delta_{\xi_1,K} + \delta_{-,\pm} \ \delta_{\xi_1,K'}\right)$, i.e., $\sigma_{+(-)}$~circularly polarized light generates interband transitions at the K$^{(\prime)}$~point.
		$\omega_0$ denotes the laser frequency.
		The second term on the right-hand side of Eq.~(\ref{eq:exciton-dyn}) couples bright excitons $P\hspace{0.5mm}^{\xi_1,s_1,s_2}_{\nu_1}$ with $s_1=s_2$ and dark excitons with $s_1 \neq s_2$ proportional to the in-plane magnetic field $B_\parallel$, cf. Fig.~\ref{Bild-AB}:
		\begin{align}
		& \partial_t \ P\hspace{0.5mm}^{\xi_1,s_1,s_2}_{\nu_1} \big|_{B_\parallel} \notag \\
		& = - \frac{\text{i} \mu_B B_\parallel}{\hbar} \sum_{\nu_2,\textbf{\textit{q}}} \varphi^*\hspace{0.5mm}^{\xi_1,s_1,\bar{s}_2}_{\nu_2,\textbf{\textit{q}}} \ \varphi\hspace{0.5mm}^{\xi_1,s_1,s_2}_{\nu_1,\textbf{\textit{q}}} \ P\hspace{0.5mm}^{\xi_1,s_1,\bar{s}_2}_{\nu_2} \notag \\
		& \hspace{4mm}
		+ \frac{\text{i} \mu_B B_\parallel}{\hbar} \sum_{\nu_2,\textbf{\textit{q}}} \varphi^*\hspace{0.5mm}^{\xi_1,\bar{s}_1,s_2}_{\nu_2,\textbf{\textit{q}}} \ \varphi\hspace{0.5mm}^{\xi_1,s_1,s_2}_{\nu_1,\textbf{\textit{q}}} \ P\hspace{0.5mm}^{\xi_1,\bar{s}_1,s_2}_{\nu_2} . \label{eq:in-plane-mixing}
		\end{align}
		The first contribution to Eq.~(\ref{eq:in-plane-mixing}) couples $P\hspace{0.5mm}^{\xi_1,s_1,s_2}_{\nu_1}$ to $P\hspace{0.5mm}^{\xi_1,s_1,\bar{s}_2}_{\nu_2}$ excitons in the same valley $\xi_1$ and with identical electron spin $s_1$ but with opposite hole spin $s_2 \neq \bar{s}_2$, i.e., $\bar{s}_2 = \ \downarrow$ for $s_2 = \ \uparrow$ and vice versa.
		The last line of Eq.~(\ref{eq:in-plane-mixing}) couples $P\hspace{0.5mm}^{\xi_1,s_1,s_2}_{\nu_1}$ to $P\hspace{0.5mm}^{\xi_1,\bar{s}_1,s_2}_{\nu_2}$ excitons in the valley $\xi_1$ with opposite electron spin $s_1 \neq \bar{s}_1$ and equal hole spin $s_2$.
		The mixing among excitons with electrons in different conduction bands and holes in the same valence band dominates (second contribution to Eq.~(\ref{eq:in-plane-mixing})).
		This is due to the small energy splitting of spin-$\uparrow$ and spin-$\downarrow$ conduction bands of a few to tens of meV compared to the significantly larger valence band splitting of more than one hundred meV \cite{kormanyos2015k}.
		Nevertheless, we take account of both terms.
		The third term on the right-hand side of Eq.~(\ref{eq:exciton-dyn}) characterizes Pauli blocking:
		\begin{align}
		& \partial_t \ P\hspace{0.5mm}^{\xi_1,s_1,s_2}_{\nu_1} \big|_\text{PB} \notag \\
		& = \text{i} \sum_{s_3,\nu_2,\nu_3} \hat{\Omega}\hspace{0.5mm}^{\xi_1,s_2,s_3,s_1}_{\nu_2,\nu_3,\nu_1} \ P\hspace{0.5mm}^{\xi_1,s_1,s_3}_{\nu_2} \left(P\hspace{0.5mm}^{\xi_1,s_1,s_3}_{\nu_3}\right)^* \notag \\
		& \hspace{4mm} + \text{i} \sum_{s_3,\nu_2,\nu_3} \hat{\Omega}\hspace{0.5mm}^{\xi_1,s_1,s_2,s_3}_{\nu_2,\nu_3,\nu_1} \ P\hspace{0.5mm}^{\xi_1,s_3,s_2}_{\nu_2} \left(P\hspace{0.5mm}^{\xi_1,s_3,s_2}_{\nu_3}\right)^* , \label{eq:Pauli-blocking}
		\end{align}
		with the Pauli blocking parameter:
		\begin{align}
			& \hat{\Omega}\hspace{0.5mm}^{\xi_1,s_1,s_2,s_3}_{\nu_2,\nu_3,\nu_1}\notag \\ %
			& = \left(\delta_{+,\pm} \ \delta_{\xi_1,K} + \delta_{-,\pm} \ \delta_{\xi_1,K'}\right) \frac{1}{\hbar}\sum_{\textbf{\textit{q}}} \varphi\hspace{0.5mm}^{\xi_1,s_1,s_2}_{\nu_1,\textbf{\textit{q}}} \notag \\
			& \hspace{4.2mm} \times \varphi^*\hspace{0.5mm}^{\xi_1,s_3,s_2}_{\nu_2,\textbf{\textit{q}}} \ \varphi\hspace{0.5mm}^{\xi_1,s_3,s_2}_{\nu_3,\textbf{\textit{q}}} \  \big(d\hspace{0.5mm}^{c,v}_{\xi_1,s_1}\big)^* \big[\tilde{E}^{\sigma_\pm}(t)\big]^* e^{\text{i} \omega_0 t}.
		\end{align}
		The first contribution to Eq.~(\ref{eq:Pauli-blocking}) originates from a coherent exciton population $\sim P\hspace{0.5mm}^{\xi_1,s_1,s_3}_{\nu_2} \left(P\hspace{0.5mm}^{\xi_1,s_1,s_3}_{\nu_3}\right)^*$ in the same valley $\xi_1$ as $P\hspace{0.5mm}^{\xi_1,s_1,s_2}_{\nu_1}$, with identical electron spin $s_1$ and either equal $s_3 = s_1$ or opposite hole spins $s_3 \neq s_1$.
		The second term on the right-hand side of Eq.~(\ref{eq:Pauli-blocking}) induces a blocking due to excitons $\sim P\hspace{0.5mm}^{\xi_1,s_3,s_2}_{\nu_2} \left(P\hspace{0.5mm}^{\xi_1,s_3,s_2}_{\nu_3}\right)^*$ with identical electron spin $s_1$ as $P\hspace{0.5mm}^{\xi_1,s_1,s_2}_{\nu_1}$ and either same $s_3 = s_2$ or opposite hole spins $s_3 \neq s_2$.
		In particular, the contributions including spin-forbidden dark excitons only appear in the presence of a magnetic field.
		The fourth term on the right-hand side of Eq.~(\ref{eq:exciton-dyn}) represents instantaneous Coulomb scattering among excitons in the same valley $\xi_1$ on a Hartree--Fock level:
		\begin{align}
		& \partial_t \ P\hspace{0.5mm}^{\xi_1,s_1,s_2}_{\nu_1} \big|_\text{HF} \notag \\
		& = \delta_{s_1,s_2} \frac{\text{i}}{\hbar} \sum_{s_3,\nu_2} W\hspace{0.5mm}^{\xi_1,s_1,s_3}_{0,\nu_2,\nu_1} \ P\hspace{0.5mm}^{\xi_1,s_3,s_3}_{\nu_2} \notag \\
		& \hspace{4mm} + \frac{\text{i}}{\hbar} \sum_{\scriptstyle s_3,s_4 \atop \scriptstyle \nu_2,\nu_3,\nu_4} W\hspace{0.5mm}^{\xi_1,s_1,s_2,s_3,s_4}_{V,\nu_2,\nu_3,\nu_4,\nu_1} \notag \\
		& \hspace{17.6mm} \times P\hspace{0.5mm}^{\xi_1,s_3,s_2}_{\nu_2} P\hspace{0.5mm}^{\xi_1,s_1,s_4}_{\nu_3} \left(P\hspace{0.5mm}^{\xi_1,s_3,s_4}_{\nu_4}\right)^* \notag \\
		& \hspace{4mm} - \frac{\text{i}}{\hbar} \sum_{\scriptstyle s_3,s_4 \atop \scriptstyle \nu_2,\nu_3,\nu_4} W\hspace{0.5mm}^{\xi_1,s_1,s_2,s_3,s_4}_{0,\nu_2,\nu_3,\nu_4,\nu_1} \notag \\
		& \hspace{21.2mm} \times P\hspace{0.5mm}^{\xi_1,s_4,s_4}_{\nu_2} P\hspace{0.5mm}^{\xi_1,s_3,s_2}_{\nu_3} \left(P\hspace{0.5mm}^{\xi_1,s_3,s_1}_{\nu_4}\right)^* \notag \\
		& \hspace{4mm} - \frac{\text{i}}{\hbar} \sum_{\scriptstyle s_3,s_4 \atop \scriptstyle \nu_2,\nu_3,\nu_4} W\hspace{0.5mm}^{\xi_1,s_2,s_3,s_1,s_4}_{0,\nu_2,\nu_3,\nu_1,\nu_4} \notag \\
		& \hspace{21.2mm} \times P\hspace{0.5mm}^{\xi_1,s_4,s_4}_{\nu_2} P\hspace{0.5mm}^{\xi_1,s_1,s_3}_{\nu_3} \left(P\hspace{0.5mm}^{\xi_1,s_2,s_3}_{\nu_4}\right)^* \notag \\
		& \hspace{4mm} + \frac{\text{i}}{\hbar} \sum_{\scriptstyle s_3,s_4 \atop \scriptstyle \nu_2,\nu_3,\nu_4} W\hspace{0.5mm}^{\xi_1,s_1,s_2,s_3,s_4}_{X,\nu_2,\nu_3,\nu_4,\nu_1} \notag \\
		& \hspace{21.2mm} \times P\hspace{0.5mm}^{\xi_1,s_3,s_2}_{\nu_2} P\hspace{0.5mm}^{\xi_1,s_4,s_3}_{\nu_3} \left(P\hspace{0.5mm}^{\xi_1,s_4,s_1}_{\nu_4}\right)^* \notag \\
		& \hspace{4mm} + \frac{\text{i}}{\hbar} \sum_{\scriptstyle s_3,s_4 \atop \nu_2,\nu_3,\nu_4} W\hspace{0.5mm}^{\xi_1,s_2,s_4,s_3,s_1}_{X,\nu_2,\nu_3,\nu_1,\nu_4} \notag \\
		& \hspace{21.2mm} \times P\hspace{0.5mm}^{\xi_1,s_3,s_4}_{\nu_2} P\hspace{0.5mm}^{\xi_1,s_1,s_3}_{\nu_3} \left(P\hspace{0.5mm}^{\xi_1,s_2,s_4}_{\nu_4}\right)^* . \label{eq:exciton-exch}
		\end{align}
		The first contribution to the right-hand side of Eq.~(\ref{eq:exciton-exch}) describes linear intravalley exchange Coulomb interactions.
		This term originates from a local field effect which only affects bright excitons with same electron and hole spin $s_1 = s_2$ \cite{qiu2015nonanalyticity,guo2019exchange}.
		This term enhances the exciton resonance energies and mixes bright A and B~excitons as well as A$'$ and B$'$~excitons.
		The matrix element $W\hspace{0.5mm}^{\xi_1,s_1,s_3}_{0,\nu_2,\nu_1}$ is given in Eq.~(\ref{eq:mat-el-exc-0}).
		All following nonlinear exciton-exciton scattering contributions on the right-hand side of Eq.~(\ref{eq:exciton-exch}) include Coulomb scattering involving not only bright but also dark excitons:
		The second term on the right-hand side of Eq.~(\ref{eq:exciton-exch}) characterizes Coulomb scattering associated with the direct Coulomb potential $W\hspace{0.5mm}^{\xi_1,s_1,s_2,s_3,s_4}_{V,\nu_2,\nu_3,\nu_4,\nu_1}$ defined in Eq.~(\ref{eq:mat-el-dir}).
		The third and fourth terms of Eq.~(\ref{eq:exciton-exch}) represent the nonlinear counterpart of the linear local field contribution (first term) with the coupling elements $W\hspace{0.5mm}^{\xi_1,s_1,s_2,s_3,s_4}_{0,\nu_2,\nu_3,\nu_4,\nu_1}$ and $W\hspace{0.5mm}^{\xi_1,s_2,s_3,s_1,s_4}_{0,\nu_2,\nu_3,\nu_1,\nu_4}$ defined in Eq.~(\ref{eq:mat-el-exc-1}).
		The last two contributions to Eq.~(\ref{eq:exciton-exch}) are associated with the exchange Coulomb matrix elements $W\hspace{0.5mm}^{\xi_1,s_1,s_2,s_3,s_4}_{X,\nu_2,\nu_3,\nu_4,\nu_1}$ and $W\hspace{0.5mm}^{\xi_1,s_2,s_4,s_3,s_1}_{X,\nu_2,\nu_3,\nu_1,\nu_4}$ defined in Eq.~(\ref{eq:mat-el-exc-2}).
		These terms originate from a $\textbf{\textit{k}}$$\cdot$$\textbf{\textit{p}}$ expansion of the exchange Coulomb potential \cite{haug2009quantum,xu2014spin,yu2014valley,wu2015exciton} resembling a dipole-dipole interaction \cite{selig2019theory}.
		The last term on the right-hand side of Eq.~(\ref{eq:exciton-exch}) incorporates exciton-exciton scattering beyond a Hartree--Fock approximation and represents the coupling of the exciton transition $P\hspace{0.5mm}^{\xi_1,s_1,s_2}_{\nu_1}$ to two-electron and two-hole transitions $B\hspace{0.5mm}^{\xi_1,s_1,s_2,\xi_2,s_3,s_4}_{\pm,\mu}$ introduced in Eq.~(\ref{eq:four-particle-corr}):
		\begin{align}
		& \partial_t \ P\hspace{0.5mm}^{\xi_1,s_1,s_2}_{\nu_1} \big|_{\text{corr.}} \notag \\
		& = \frac{\text{i}}{\hbar} \sum_{\scriptstyle \xi_2,s_3,s_4 \atop \scriptstyle \nu_2,\nu_3,\nu_4,\pm,\textbf{\textit{P}}} \left(\hat{W}\hspace{0.5mm}^{\xi_1,s_2,\xi_2,s_4}_{\pm,\nu_2,\nu_3,\nu_1,\nu_4,\textbf{\textit{P}},\textbf{0}}\right)^* \left(P\hspace{0.5mm}^{\xi_2,s_3,s_4}_{\nu_4}\right)^* \notag \\
		& \hspace{27.4mm} \times \sum_{\mu} \Phi^{\text{R}}\hspace{0.5mm}^{\xi_1,s_2,\xi_2,s_4}_{\pm,\mu,\nu_2,\nu_3,\textbf{\textit{P}}} \ B\hspace{0.5mm}^{\xi_1,s_1,s_2,\xi_2,s_3,s_4}_{\pm,\mu} \notag \\
		& \hspace{4mm} - \frac{\text{i}}{\hbar} \sum_{ \scriptstyle \xi_2,s_3,s_4 \atop \scriptstyle \nu_2,\nu_3,\nu_4,\pm,\textbf{\textit{P}}} \left(\hat{X}\hspace{0.5mm}^{\xi_1,s_1,s_2,\xi_2,s_4}_{\pm,\nu_2,\nu_3,\nu_1,\nu_4,\textbf{\textit{P}}}\right)^* \left(P\hspace{0.5mm}^{\xi_1,s_3,s_1}_{\nu_4}\right)^* \notag \\
		& \hspace{27.4mm} \times \sum_{\mu} \Phi^{\text{R}}\hspace{0.5mm}^{\xi_1,s_2,\xi_2,s_4}_{\pm,\mu,\nu_2,\nu_3,\textbf{\textit{P}}} \ B\hspace{0.5mm}^{\xi_1,s_3,s_2,\xi_2,s_4,s_4}_{\pm,\mu} \notag \\
		& \hspace{4mm} - \frac{\text{i}}{\hbar} \sum_{\scriptstyle \xi_2,s_3,s_4 \atop \scriptstyle \nu_2,\nu_3,\nu_4,\pm,\textbf{\textit{P}}} \left(\hat{X}\hspace{0.5mm}^{\xi_1,s_2,s_3,\xi_2,s_4}_{\pm,\nu_2,\nu_3,\nu_4,\nu_1,\textbf{\textit{P}}}\right)^* \left(P\hspace{0.5mm}^{\xi_1,s_2,s_3}_{\nu_4}\right)^* \notag \\
		& \hspace{27.4mm} \times \sum_{\mu} \Phi^{\text{R}}\hspace{0.5mm}^{\xi_1,s_3,\xi_2,s_4}_{\pm,\mu,\nu_2,\nu_3,\textbf{\textit{P}}} \ B\hspace{0.5mm}^{\xi_1,s_1,s_3,\xi_2,s_4,s_4}_{\pm,\mu} . \label{eq:corr}
		\end{align}
		Direct Coulomb scattering, associated with the Coulomb matrix $\hat{W}\hspace{0.5mm}^{\xi_1,s_2,\xi_2,s_4}_{\pm,\nu_2,\nu_3,\nu_1,\nu_4,\textbf{\textit{P}},\textbf{0}}$ given in Eq.~(\ref{eq:mat-el-bix-dir}), couples the exciton transition $P\hspace{0.5mm}^{\xi_1,s_1,s_2}_{\nu_1}$ to the exciton transition and two-electron and two-hole Coulomb correlations $\left(P\hspace{0.5mm}^{\xi_2,s_3,s_4}_{\nu_4}\right)^* B\hspace{0.5mm}^{\xi_1,s_1,s_2,\xi_2,s_3,s_4}_{\pm,\mu}$.
		Exchange Coulomb interaction couples the exciton transition $P\hspace{0.5mm}^{\xi_1,s_1,s_2}_{\nu_1}$ to  $\left(P\hspace{0.5mm}^{\xi_1,s_3,s_1}_{\nu_4}\right)^*B\hspace{0.5mm}^{\xi_1,s_3,s_2,\xi_2,s_4,s_4}_{\pm,\mu}$ accompanied by the Coulomb matrix $\hat{X}\hspace{0.5mm}^{\xi_1,s_1,s_2,\xi_2,s_4}_{\pm,\nu_2,\nu_3,\nu_1,\nu_4,\textbf{\textit{P}}}$ as well as to $\left(P\hspace{0.5mm}^{\xi_1,s_2,s_3}_{\nu_4}\right)^* B\hspace{0.5mm}^{\xi_1,s_1,s_3,\xi_2,s_4,s_4}_{\pm,\mu}$ via $\hat{X}\hspace{0.5mm}^{\xi_1,s_2,s_3,\xi_2,s_4}_{\pm,\nu_2,\nu_3,\nu_4,\nu_1,\textbf{\textit{P}}}$ defined in Eq.~(\ref{eq:mat-el-bix-exc}).
		In contrast to nonlinear exciton-exciton interaction on a Hartree--Fock level, Eq.~(\ref{eq:corr}) includes not only intravalley scattering $\xi_2 = \xi_1$ but also intervalley scattering $\xi_2 \neq \xi_1 $.
		The two-electron and two-hole Coulomb correlation dynamics for  $B\hspace{0.5mm}^{\xi_1,s_1,s_2,\xi_2,s_3,s_4}_{\pm,\mu}$ is described by:
		\begin{align}
		& \bigg(\partial_t +\gamma\hspace{0.5mm}^{\xi_1,s_1,s_2}_{\text{x}} + \gamma\hspace{0.5mm}^{\xi_2,s_3,s_4}_{\text{x}} \notag \\
		& \hspace{2mm} - \frac{\text{i}}{\hbar} \hat{\epsilon}\hspace{0.5mm}^{\xi_1,s_1,s_2,\xi_2,s_3,s_4}_{\text{xx},\pm,\mu,B_\perp,B_\parallel} \bigg) B\hspace{0.5mm}^{\xi_1,s_1,s_2,\xi_2,s_3,s_4}_{\pm,\mu} \notag \\
		%
		%
		& = \frac{\text{i}\mu_B B_\parallel}{\hbar} \sum_{\scriptstyle \nu_1,\dots,\nu_6 \atop \scriptstyle \mu',\textbf{\textit{P}}, \textbf{\textit{k}},\textbf{\textit{K}}} \Phi^{\text{L}}\hspace{0.5mm}^{\xi_1,s_2,\xi_2,s_4}_{\pm,\mu,\nu_1,\nu_2,\textbf{\textit{P}}} \big(S\hspace{0.5mm}^{\xi_1,s_2,\xi_2,s_4}_{\pm}\big)^{-1}_{\nu_1,\nu_2,\nu_3,\nu_4,\textbf{\textit{P}},\textbf{\textit{k}}} \notag \\
		& \hspace{3.6mm} \times \Big[ S\hspace{0.5mm}^{\xi_1,s_2,\xi_2,s_4}_{\pm,\nu_3,\nu_4,\nu_5,\nu_6,\textbf{\textit{k}},\textbf{\textit{K}}} \Phi^{\text{R}}\hspace{0.5mm}^{\xi_1,s_2,\xi_2,s_4}_{\pm,\mu',\nu_5,\nu_6,\textbf{\textit{K}}} \notag \\
		& \hspace{9.6mm} \times \big(B\hspace{0.5mm}^{\xi_1,\bar{s}_1,s_2,\xi_2,s_3,s_4}_{\pm,\mu'} + B\hspace{0.5mm}^{\xi_1,s_1,s_2,\xi_2,\bar{s}_3,s_4}_{\pm,\mu'}\big) \notag \\
		& \hspace{9.6mm} - S\hspace{0.5mm}^{\xi_1,\bar{s}_2,\xi_2,s_4}_{\pm,\nu_3,\nu_4,\nu_5,\nu_6,\textbf{\textit{k}},\textbf{\textit{K}}} \Phi^{\text{R}}\hspace{0.5mm}^{\xi_1,\bar{s}_2,\xi_2,s_4}_{\pm,\mu',\nu_5,\nu_6,\textbf{\textit{K}}} B\hspace{0.5mm}^{\xi_1,s_1,\bar{s}_2,\xi_2,s_3,s_4}_{\pm,\mu'} \notag \\
		& \hspace{9.6mm} - S\hspace{0.5mm}^{\xi_1,s_2,\xi_2,\bar{s}_4}_{\pm,\nu_3,\nu_4,\nu_5,\nu_6,\textbf{\textit{k}},\textbf{\textit{K}}} \Phi^{\text{R}}\hspace{0.5mm}^{\xi_1,s_2,\xi_2,\bar{s}_4}_{\pm,\mu',\nu_5,\nu_6,\textbf{\textit{K}}} B\hspace{0.5mm}^{\xi_1,s_1,s_2,\xi_2,s_3,\bar{s}_4}_{\pm,\mu'} \Big] \notag \\
\end{align}
\begin{align}
		& \hspace{4mm} + \frac{\text{i}}{2\hbar} \sum_{\scriptstyle \nu_1,\dots,\nu_6 \atop \scriptstyle \textbf{\textit{P}},\textbf{\textit{k}}} \Phi^{\text{L}}\hspace{0.5mm}^{\xi_1,s_2,\xi_2,s_4}_{\pm,\mu,\nu_1,\nu_2,\textbf{\textit{P}}} \big(S\hspace{0.5mm}^{\xi_1,s_2,\xi_2,s_4}_{\pm}\big)^{-1}_{\nu_1,\nu_2,\nu_3,\nu_4,\textbf{\textit{P}},\textbf{\textit{k}}} \notag \\
		& \hspace{7.6mm} \times \Big[ \hat{W}\hspace{0.5mm}^{\xi_1,s_2,\xi_2,s_4}_{\pm,\nu_3,\nu_4,\nu_5,\nu_6,\textbf{\textit{k}},\textbf{0}} \big(P\hspace{0.5mm}^{\xi_1,s_1,s_2}_{\nu_5} P\hspace{0.5mm}^{\xi_2,s_3,s_4}_{\nu_6} \notag \\
		& \hspace{40.7mm}\pm \delta_{\xi_1,\xi_2} P\hspace{0.5mm}^{\xi_1,s_3,s_2}_{\nu_5} P\hspace{0.5mm}^{\xi_1,s_1,s_4}_{\nu_6} \big) \notag \\
		& \hspace{12.6mm} - \sum_{s_5} \delta_{s_3,s_4}  \hat{X}\hspace{0.5mm}^{\xi_1,s_5,s_2,\xi_2,s_3}_{\pm,\nu_3,\nu_4,\nu_5,\nu_6,\textbf{\textit{k}}} P\hspace{0.5mm}^{\xi_1,s_5,s_2}_{\nu_5} P\hspace{0.5mm}^{\xi_1,s_1,s_5}_{\nu_6} \notag \\
		& \hspace{12.6mm} \mp \sum_{s_5} \delta^{\xi_1,\xi_2}_{s_1,s_4} \hat{X}\hspace{0.5mm}^{\xi_1,s_5,s_2,\xi_2,s_1}_{\pm,\nu_3,\nu_4,\nu_5,\nu_6,\textbf{\textit{k}}} P\hspace{0.5mm}^{\xi_1,s_5,s_2}_{\nu_5} P\hspace{0.5mm}^{\xi_1,s_3,s_5}_{\nu_6} \notag \\
		& \hspace{12.6mm} - \sum_{s_5}  \delta_{s_1,s_2} \hat{X}\hspace{0.5mm}^{\xi_2,s_5,s_4,\xi_1,s_1}_{\pm,\nu_4,\nu_3,\nu_6,\nu_5,-\textbf{\textit{k}}} P\hspace{0.5mm}^{\xi_2,s_3,s_5}_{\nu_5} P\hspace{0.5mm}^{\xi_2,s_5,s_4}_{\nu_6} \notag \\
		& \hspace{12.6mm} \mp \sum_{s_5} \delta^{\xi_1,\xi_2}_{s_2,s_3} \hat{X}\hspace{0.5mm}^{\xi_2,s_5,s_4,\xi_1,s_2}_{\pm,\nu_4,\nu_3,\nu_6,\nu_5,-\textbf{\textit{k}}} P\hspace{0.5mm}^{\xi_2,s_1,s_5}_{\nu_5} P\hspace{0.5mm}^{\xi_2,s_5,s_4}_{\nu_6} \Big]
		. \label{eq:bix-dyn}
		\end{align}
		The left-hand side of Eq.~(\ref{eq:bix-dyn}) describes oscillations with energy $\hat{\epsilon}\hspace{0.5mm}^{\xi_1,s_1,s_2,\xi_2,s_3,s_4}_{\text{xx},\pm,\mu,B_\perp,B_\parallel}$damped by $\gamma\hspace{0.5mm}^{\xi_1,s_1,s_2}_{\text{x}} + \gamma\hspace{0.5mm}^{\xi_2,s_3,s_4}_{\text{x}}$.
		The resonance energy $\hat{\epsilon}\hspace{0.5mm}^{\xi_1,s_1,s_2,\xi_2,s_3,s_4}_{\text{xx},\pm,\mu,B_\perp,B_\parallel}$ involves the two-electron and two-hole correlation energy $\epsilon\hspace{0.5mm}^{\xi_1,s_1,s_2,\xi_2,s_3,s_4}_{\text{xx},\pm,\mu}$ obtained from solving the two-electron and two-hole Schrödinger equation which are renormalized by $\Delta\hspace{0.5mm}^{\xi_1,s_1,s_2}_{\text{x},B_\perp,B_\parallel}$ and $\Delta\hspace{0.5mm}^{\xi_2,s_3,s_4}_{\text{x},B_\perp,B_\parallel}$:
		\begin{align}
		\hat{\epsilon}\hspace{0.5mm}^{\xi_1,s_1,s_2,\xi_2,s_3,s_4}_{\text{xx},\pm,\mu,B_\perp} = \ & \epsilon\hspace{0.5mm}^{\xi_1,s_1,s_2,\xi_2,s_3,s_4}_{\text{xx},\pm,\mu} \notag \\
		& + \Delta\hspace{0.5mm}^{\xi_1,s_1,s_2}_{\text{x},B_\perp,B_\parallel} + \Delta\hspace{0.5mm}^{\xi_2,s_3,s_4}_{\text{x},B_\perp,B_\parallel}.
		\end{align}
		The renormalization $\Delta\hspace{0.5mm}^{\xi_1,s_1,s_2}_{\text{x},B_\perp,B_\parallel}$ is obtained by firstly, diagonalization of the eight-dimensional linear exciton Hamiltonian spanned by $P\hspace{0.5mm}^{\xi_1,s_1,s_2}_{\text{1s}}$ ($\xi_1 = \ $K, K$'$, $s_1 = \ \uparrow, \downarrow$, $s_2 = \ \uparrow, \downarrow$) including $B_\perp$ and $B_\parallel$ dependent energy renormalizations (Eqs.~(\ref{eq:Zeeman-1}), (\ref{eq:Zeeman-2}), and (\ref{eq:in-plane-mixing})) and linear local field exchange Coulomb scattering (first term  of Eq.~(\ref{eq:exciton-exch})), and secondly, subtracting the respective exciton binding energies $\epsilon\hspace{0.5mm}^{\xi_1,s_1,s_2}_{\text{x},\text{1s}}$ determined by the Wannier equation.
		The first term on the right-hand side of Eq.~(\ref{eq:bix-dyn}) characterizes the mixing among two-electron and two-hole Coulomb correlations proportional to the in-plane magnetic field $B_\parallel$:
		$B\hspace{0.5mm}^{\xi_1,s_1,s_2,\xi_2,s_3,s_4}_{\pm,\mu}$ couples to correlations with opposite electron spin $s_1 \neq \bar{s}_1$ ($B\hspace{0.5mm}^{\xi_1,\bar{s}_1,s_2,\xi_2,s_3,s_4}_{\pm,\mu'}$) and $s_3 \neq \bar{s}_3$ ($B\hspace{0.5mm}^{\xi_1,s_1,s_2,\xi_2,\bar{s}_3,s_4}_{\pm,\mu'}$) as well as to correlations with different hole spin $s_2 \neq \bar{s}_2$ ($B\hspace{0.5mm}^{\xi_1,s_1,\bar{s}_2,\xi_2,s_3,s_4}_{\pm,\mu'}$) or $s_4 \neq \bar{s}_4$ ($B\hspace{0.5mm}^{\xi_1,s_1,s_2,\xi_2,s_3,\bar{s}_4}_{\pm,\mu'}$).
		The matrix $S\hspace{0.5mm}^{\xi_1,s_2,\xi_2,s_4}_{\pm,\nu_1,\nu_2,\nu_3,\nu_4,\textbf{\textit{P}},\textbf{\textit{k}}}$ directly follows from the definition of the two-electron and two-hole correlation function in Eq.~(\ref{eq:four-particle-corr}) and is defined in Eq.~(\ref{eq:S-Matrix}).
		Since $S\hspace{0.5mm}^{\xi_1,s_2,\xi_2,s_4}_{\pm,\nu_1,\nu_2,\nu_3,\nu_4,\textbf{\textit{P}},\textbf{\textit{k}}}$ is solely determined by the conduction and valence band curvatures, which are very similar for monolayer TMDCs \cite{kormanyos2015k}, the coupling matrix is approximated by:
		\begin{align}
		& \sum_{\nu_1,\nu_2,\textbf{\textit{P}}} \Phi^{\text{L}} \hspace{0.5mm}^{\xi_1,s_2,\xi_2,s_4}_{\pm,\mu,\nu_1,\nu_2,\textbf{\textit{P}}} \sum_{\nu_3,\nu_4,\textbf{\textit{k}}} \left(S\hspace{0.5mm}^{\xi_1,s_2,\xi_2,s_4}_{\pm}\right)^{-1}_{\nu_1,\nu_2,\nu_3,\nu_4,\textbf{\textit{P}},\textbf{\textit{k}}} \notag \\
		& \times \sum_{\nu_5,\nu_6,\textbf{\textit{K}}} S\hspace{0.5mm}^{\xi_1,s_2,\xi_2,s_4}_{\pm,\nu_3,\nu_4,\nu_5,\nu_6,\textbf{\textit{k}},\textbf{\textit{K}}} \ \Phi^{\text{R}}\hspace{0.5mm}^{\xi_1,s_2,\xi_2,s_4}_{\pm,\mu',\nu_5,\nu_6,\textbf{\textit{K}}} \notag \\
		& \approx \delta_{\mu,\mu'} .
		\end{align}
		The second contribution to the right-hand side of Eq.~(\ref{eq:bix-dyn}) describes Coulomb-driven source terms of $B\hspace{0.5mm}^{\xi_1,s_1,s_2,\xi_2,s_3,s_4}_{\pm,\mu}$ due to two exciton transitions $P\hspace{0.5mm}^{\xi_1,s_1,s_2}_{\nu_5} P\hspace{0.5mm}^{\xi_2,s_3,s_4}_{\nu_6}$.
		In particular, an in-plane magnetic field does not only account for source terms due to bright excitons ($s_1=s_2$ and $s_3=s_4$) but includes dark exciton source terms ($s_1\neq s_2$ or $s_3\neq s_4$) as well.
		The appearing direct $\hat{W}\hspace{0.5mm}^{\xi_1,s_2,\xi_2,s_4}_{\pm,\nu_3,\nu_4,\nu_5,\nu_6,\textbf{\textit{k}},\textbf{0}}$ and exchange Coulomb matrices $\hat{X}\hspace{0.5mm}^{\xi_1,s_5,s_2,\xi_2,s_3}_{\pm,\nu_3,\nu_4,\nu_5,\nu_6,\textbf{\textit{k}}}$ are defined in Eqs.~(\ref{eq:mat-el-bix-dir}) and (\ref{eq:mat-el-bix-exc}), respectively. \\

\section{Results} \label{sec:results}

		A careful investigation of our equations of motion in section~\ref{app:eqs-of-motion} reveals that a multitude of new effects appear in the presence of magnetic fields, which have to be considered in the interpretation of experiments:
		\begin{itemize}
			\setlength{\parskip}{0pt}
			\setlength{\parsep}{0pt}
			
			\item[(1)] out-of-plane magnetic-field-dependent Zeeman shifts of the exciton energies,
			
			\item[(2)] in-plane magnetic-field-dependent brightening of dark excitons $P\hspace{0.5mm}^{\xi,s_1,s_2}_{\nu}$ (opposite electron and hole spins $s_1 \neq s_2$) which couple to bright excitons $P\hspace{0.5mm}^{\xi,s_1,s_1}_{\nu}$ (same electron and hole spins $s_1 = s_2$),
			
			\item[(3)] additional Pauli blocking contributions associated with coherent dark exciton densities,
			
			\item[(4)] new direct and exchange exciton-exciton scattering terms accounting for Coulomb interactions including dark excitons,
			
			\item[(5)] further two-electron and two-hole Coulomb correlations $B\hspace{0.5mm}^{\xi_1,s_1,s_2,\xi_2,s_3,s_4}_{\pm,\mu}$ with $s_1 \neq s_2$ or $s_3 \neq s_4$ representing new Coulomb scattering channels,
			
			\item[(6)] out-of-plane magnetic-field-dependent Zeeman shifts of the two-electron and two-hole correlation energies,
			
			\item[(7)] in-plane magnetic-field-dependent coupling among two-electron and two-hole Coulomb correlations, and
			
			\item[(8)] additional source terms of the two-electron and two-hole Coulomb correlations due to dark excitons.
			
		\end{itemize}
		The coupled dynamics of exciton transitions and two-electron and two-hole Coulomb correlations described by the excitonic Bloch equations, Eqs.~(\ref{eq:exciton-dyn}) and (\ref{eq:bix-dyn}), are numerically evaluated together with Maxwell's wave equation \cite{stroucken1996coherent,knorr1996theory} for the energetically lowest $\nu =$ 1s, 2s, and 3s exciton transitions and the corresponding two-electron and two-hole correlations with s-symmetry of MoS$_2$ encapsulated in hBN at a temperature of 5~K.
		The required material parameters are summarized in Ref.~\cite{katsch2020theory}.
		In the following, the linear transmission and nonlinear differential transmission of monolayer MoS$_2$ encapsulated in hBN are separately discussed for different magnetic field orientations with respect to the monolayer plane:
		an out-of-plane magnetic field $B_\perp$ (subsection~\ref{sub:out}), an in-plane magnetic field $B_\parallel$ (subsection~\ref{sub:in}), and a magnetic field $B_\angle$ oriented in a tilt angle of 45° (subsection~\ref{sub:tilted}).
		For the nonlinear transmission we choose a $\sigma_+$~circularly polarized 50~fs Gaussian pump pulse (intensity FWHM).
		Its access energy is resonant to the respective magnetic-field-dependent A$_\text{1s}$~exciton energy.
		The differential transmission spectrum (DTS) $\delta T(\omega) = T_{\text{p}+\text{t}}(\omega) - T_{\text{t}}(\omega)$ of the probe pulse is defined as the transmission of the pumped system $T_{\text{p}+\text{t}}(\omega)$ minus the linear transmission of the probe pulse $T_{\text{t}}(\omega)$.
		In order to directly visualize the pump-induced changes of the transmission, we do not divide the DTS by the linear transmission of the test pulse $T_{\text{t}}(\omega)$.
		However, dividing by $T_{\text{t}}(\omega)$ would not qualitatively change the DTS but only enhance the signal directly at the exciton resonances while features which are further away from the exciton resonances are less pronounced.
		The energetically broadband 1~fs probe pulse is either $\sigma_+$~circularly polarized to investigate intravalley exciton-exciton interaction or $\sigma_-$~circularly polarized in order to study intervalley scattering, cf. optical selection rules as indicated in Fig.~\ref{Bild-AB}.
		Assuming zero time delay between pump and probe pulses allows to neglect contributions from incoherent exciton densities in the following \cite{selig2019ultrafast,selig2019quenching,christiansen2019theory}.

\subsection{Out-of-plane Magnetic Field} \label{sub:out}

		\begin{figure*}
			\centering
			\includegraphics[width=1.95\columnwidth]{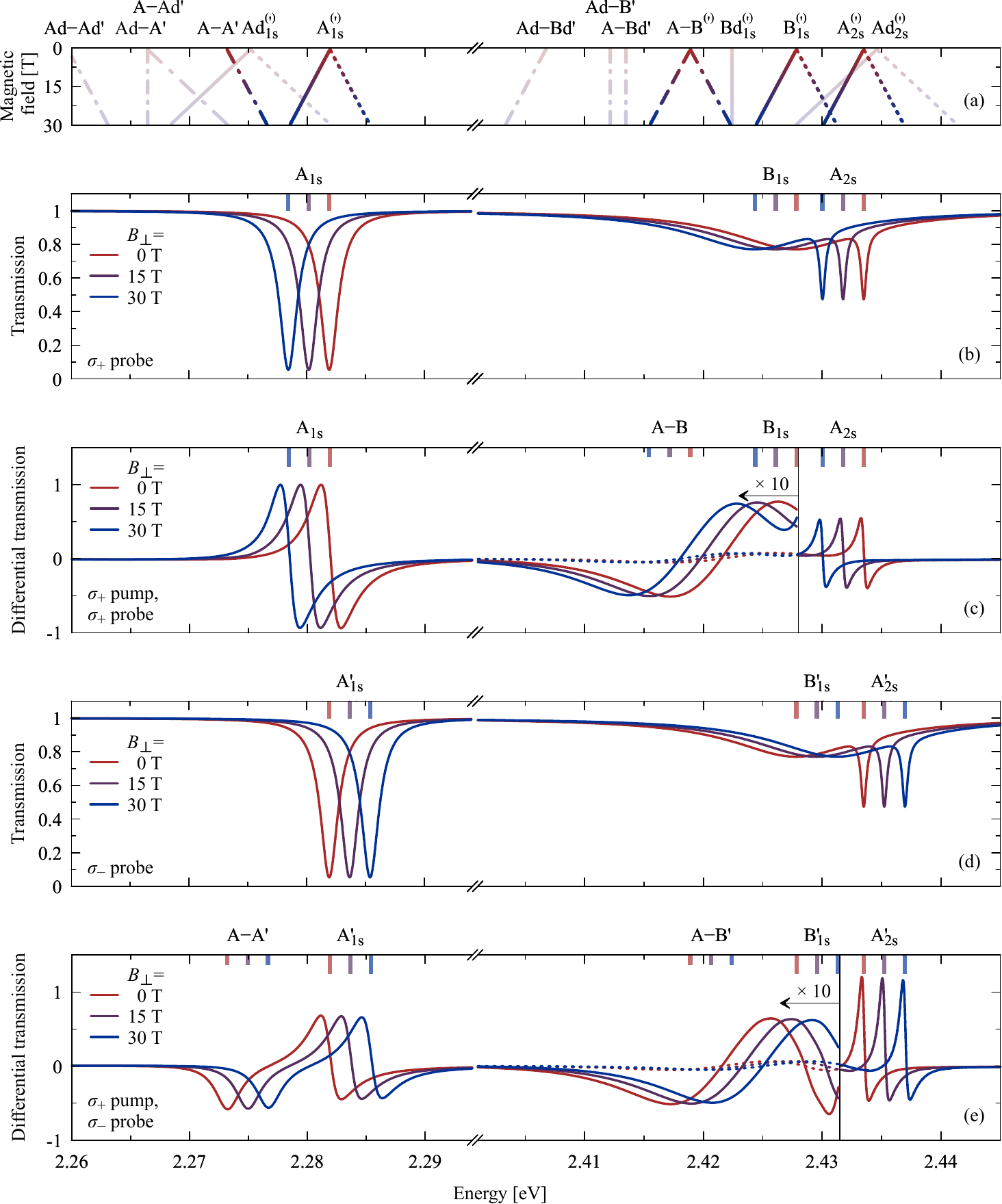}
			\caption{\textbf{Out-of-plane magnetic-field-dependent transmission and differential transmission.}
				(a)~Exciton and biexciton resonance energies for different out-of-plane magnetic fields $B_\perp$, cf. Figs.~\ref{Bild-AB} and \ref{fig:bix-schema}.
				Solid (dotted) lines represent exciton resonances and dashed (dash-dotted) lines biexciton resonances for $\sigma_{+}$ ($\sigma_-$) circularly polarized excitation.
				Pale lines indicate vanishing oscillator strengths of the resonances.
				(b,d)~Linear transmission spectra of monolayer MoS$_2$ encapsulated in hBN at 5~K for (b)~$\sigma_+$ and (d)~$\sigma_-$~circularly polarized light for different $B_\perp$.
				The $B_\perp$ dependent exciton energies are indicated as colored bars above.
				(c,e)~Normalized differential transmission spectra for different magnetic fields $B_\perp$.
				The A$_\text{1s}$~exciton is resonantly pumped by $\sigma_+$~circularly polarized light and the energetically broadband probe pulse is either (c)~$\sigma_+$~circularly polarized or (e)~$\sigma_-$~circularly polarized.
				The differential transmission is partially enhanced as indicated while the original signal is plotted by dotted lines.
				The colored long bars above mark the exciton energies, whereas the shorter bars indicate the biexciton energies.
			}
			\label{fig:out-of-plane-dts}
		\end{figure*}
		\textit{Linear response:}
		With increasing out-of-plane magnetic field $B_\perp>0$~T, the A and B~excitons associated with the K~valley experience Zeeman shifts toward lower energies shown as solid lines in Fig.~\ref{fig:out-of-plane-dts}\,(a).
		Simultaneously, the A$'$ and B$'$~transitions plotted as dotted lines in Fig.~\ref{fig:out-of-plane-dts}\,(a) shift toward higher energies due to Zeeman shifts with opposite sign for the K$'$~valley.
		The linear transmission spectrum of monolayer MoS$_2$ at zero magnetic field $B_\perp=0$~T is plotted as red lines in Fig.~\ref{fig:out-of-plane-dts}~(b,d) and shows two prominent exciton resonances which are referred to as A$_\text{1s}$ and B$_\text{1s}$~excitons with the exciton quantum number $\nu = \text{1s}$.
		Additionally, the A$_\text{2s}$ exciton resonance associated with the quantum number $\nu = \text{2s}$ of the A~series can be observed energetically above the B$_\text{1s}$ exciton.
		As a benchmark to recent literature, for instance Ref.~\cite{van2018strong}, the out-of-plane magnetic field $B_\perp$ dependent linear transmission is shown in Fig.~\ref{fig:out-of-plane-dts}\,(b,d) for $\sigma_+$ and $\sigma_-$~circularly polarized light, respectively.
		The red curves in Fig.~\ref{fig:out-of-plane-dts}\,(b,d) represent the linear transmission at $B_\perp=0$~T which is identical for $\sigma_+$ and $\sigma_-$~circular excitation.
		In contrast, the spectra at $B_\perp=15$~T and $B_\perp=30$~T, plotted as purple and blue lines in Fig.~\ref{fig:out-of-plane-dts}\,(b,d), mirror the opposite Zeeman shifts of A, B and A$'$, B$'$~excitons.
		Our microscopically calculated A$_\text{1s}$~exciton linewidth includes a phonon-mediated part of $2\gamma\hspace{0.5mm}^{\text{A}_\text{1s}}_{\text{x}} = 0.4$~meV at 5~K and a radiative part of $2\gamma\hspace{0.5mm}^{\text{A}_\text{1s}}_{\text{r}} = 1.4$~meV.
		The B$_\text{1s}$~exciton linewidth exhibits a much larger phonon-induced contribution of $2\gamma\hspace{0.5mm}^{\text{B}_\text{1s}}_{\text{x}} = 11.6$~meV at 5~K which overshadows the radiative part of $2\gamma\hspace{0.5mm}^{\text{B}_\text{1s}}_{\text{r}} = 1.6$~meV.
		The increased phonon-mediated  B$_\text{1s}$~linewidth contribution $\gamma\hspace{0.5mm}^{\text{B}_\text{1s}}_{\text{x}}$ stems from pronounced emission of acoustic and optical K~phonons which drive the relaxation of B$_\text{1s}$~excitons into exciton states including a hole at the $\Gamma$~point \cite{khatibi2018impact}.
		In contrast, the reduced A$_\text{2s}$~linewidth originates from a reduced radiative linewidth contribution of only $2\gamma\hspace{0.5mm}^{\text{A}_\text{2s}}_{\text{r}} = 0.2$~meV due to the increased spatial extent of exciton wave functions with higher exciton quantum numbers \cite{brem2019intrinsic,boule2020coherent}.
		In particular, the low A$_\text{2s}$~linewidth for high quality TMDCs leads to a relatively large oscillator strength.
		\begin{figure}
			\centering
			\includegraphics[width=0.8\columnwidth]{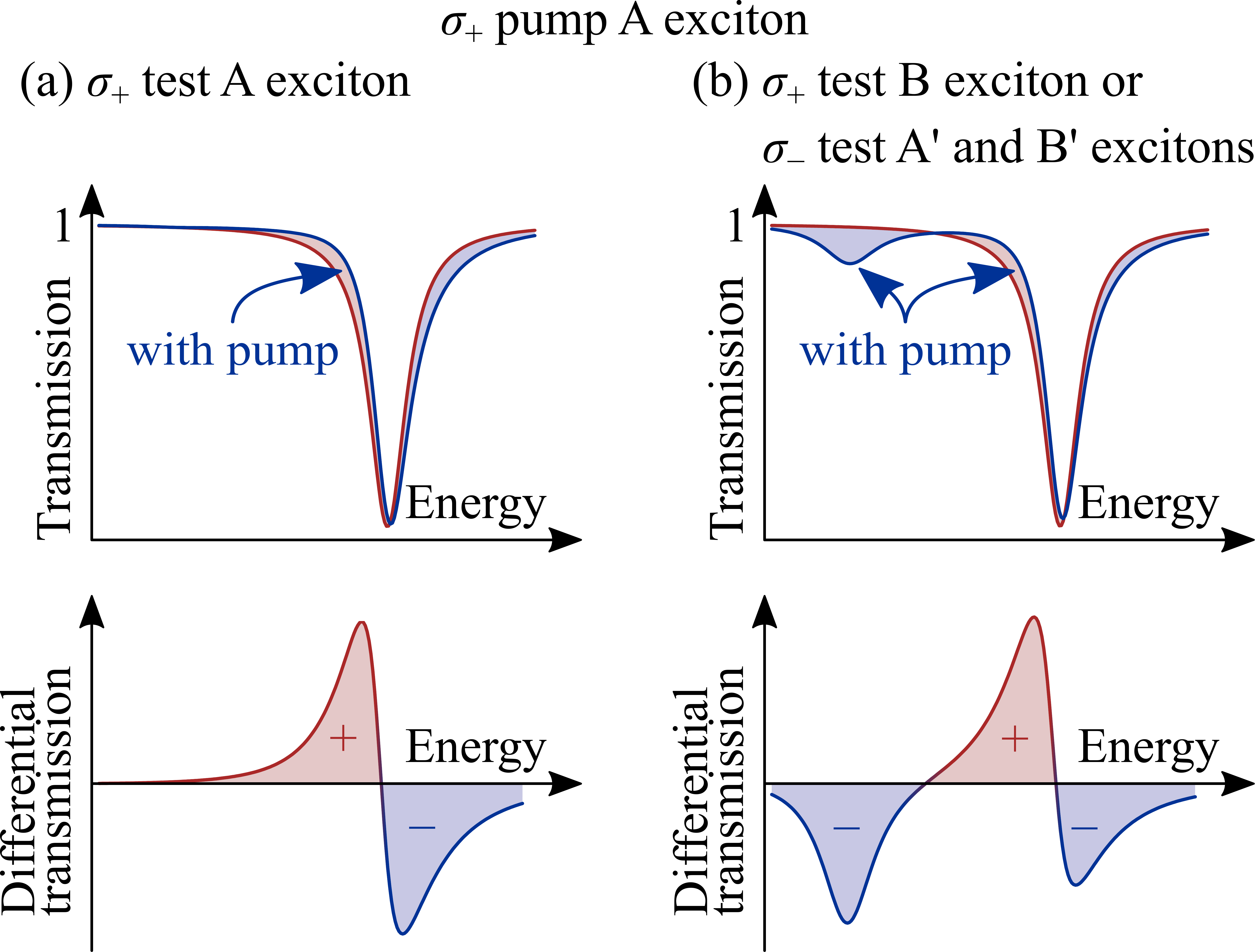}
			\caption{\textbf{Emergence of excitonic differential transmission signatures.}
				Illustrated is the transmission with (red) and without (blue) a $\sigma_+$~circularly polarized pump pulse resonant to the A~exciton near an excitonic resonance as indicated.
				(a)~A $\sigma_+$~circularly polarized test pulse gives a blue shift as well as an exciton-scattering-induced sideband on the high energy side of the A~exciton energy.
				This induces differential transmission with first a positive and than a negative contribution.
				(b)~For a $\sigma_+$ or $\sigma_-$~circularly polarized test pulse the differential transmission near the B or A$'$, B$'$~excitons exhibits first a negative feature from a biexciton resonance, followed by a positive and than negative contribution from the blue shifted exciton with an exciton-scattering-induced sideband on the high energy side.
			}
			\label{dts-schema}
		\end{figure}
		\textit{DTS for $\sigma_+$ pump and $\sigma_+$ probe pulses:}
		At first, we recapitulate the expected DTS for a vanishing magnetic field $B_\perp = 0$~T plotted as red curve in Fig.~\ref{fig:out-of-plane-dts}\,(c).
		Here, the A~excitons shifts blue giving a dispersive DTS signal with a positive contribution followed by a  negative one as schematically illustrated in Fig.~\ref{dts-schema}\,(a).
		Asymmetric sidebands on the high energy side of the A~excitons, originating from the intravalley A$-$A exciton-exciton scattering continuum, further enhance the negative DTS contributions \cite{katsch2020exciton}.
		In contrast, the DTS near the B~resonances is expected to first show a negative DTS signal below the B~exciton energy as illustrated in Fig.~\ref{dts-schema}\,(b) which corresponds to the intravalley A$-$B~biexciton resonance (Fig.~\ref{fig:bix-schema}\,(a)).
		Moreover, a blue shift of the B~resonances together with exciton-scattering-induced sidebands (due to the A$-$B~exciton-exciton scattering continuum) \cite{katsch2020exciton} induce a dispersive DTS signal with a positive feature followed by a negative contribution above the B$_\text{1s}$~exciton energy.
		The increased B$_\text{1s}$~exciton linewidth compared to A$_\text{1s}$~excitons results in weaker DTS signals since the oscillator strength is distributed over a larger energy range.
		Therefore, the energy range around the B$_\text{1s}$~exciton is shown ten times enhanced as indicated by the arrow in Fig.~\ref{fig:out-of-plane-dts}\,(c).
		Additionally, the expected negative DTS feature above the B$_\text{1s}$~excitons is absent due to the positive DTS contribution of A$_\text{2s}$~excitons.
		The DTS for an out-of-plane magnetic field of $B_\perp = 15$~T and $B_\perp = 30$~T are shown as purple and blue lines in Fig.~\ref{fig:out-of-plane-dts}\,(c).
		With increasing $B_\perp$ the whole DTS shifts towards lower energies according to the Zeeman shifted A and B~exciton resonances.
		Interestingly this holds also true for the intravalley A$-$B intravalley biexciton resonance, such that the energetic position of the biexciton resonance with respect to the B$_\text{1s}$~exciton is unchanged due to the same Zeeman shifts, cf. also dashed line in Fig.~\ref{fig:out-of-plane-dts}\,(a).
		Even though a simple band analysis suggests a doubled biexciton $g$-factor with respect to the B$_\text{1s}$ exciton, this expectation is not mirrored in the differential transmission.
		While the biexciton $B\hspace{0.5mm}^{\text{A\smash[b]{,}B}}_{-,\mu=b}$ described by Eq.~(\ref{eq:bix-dyn}) indeed has a doubled $g$-factor, only the biexciton resonance $B\hspace{0.5mm}^{\text{A\smash[b]{,}B}}_{-,\mu=b}(P\hspace{0.5mm}^{\text{A}}_{\text{1s}})^*$ couples to the exciton resonance determined by Eq.~(\ref{eq:corr}).
		As the $g$-factors of the biexciton resonance $B\hspace{0.5mm}^{\text{A\smash[b]{,}B}}_{-,\mu=b}(P\hspace{0.5mm}^{\text{A}}_{\text{1s}})^*$ associated with the A$_\text{1s}$ exciton cancel, the biexciton resonance inherits the $g$-factor of the B$_\text{1s}$ exciton.
		Negative magnetic fields $B_\perp<0$~T lead to DTS shifted in the opposite energetic direction, cf. Appendix~\ref{sec:app-dts}.
		\textit{DTS for $\sigma_+$ pump and $\sigma_-$ probe pulses:}
		At first, we again discuss the expected DTS at zero magnetic fields $B_\perp = 0$~T plotted as a red curve in Fig.~\ref{fig:out-of-plane-dts}\,(e).
		Here, the intervalley A$-$A$'$ and A$-$B$'$~biexciton resonances (Fig.~\ref{fig:bix-schema}\,(b) and (f)) lead to negative signatures in the DTS below the A$'$ and B$'$~energies.
		These negative contributions follow first a positive and then a negative DTS signal accounting for the blue shifted A$'$ and B$'$~transitions with exciton-scattering-induced sidebands on the high energy sides of the exciton resonances \cite{katsch2020exciton}.
		The latter originate from the A$-$A$'$~and A$-$B$'$~exciton-exciton scattering continua.
		The resulting DTS is schematically illustrated in Fig.~\ref{dts-schema}\,(b).
		Applying an out-of-plane magnetic field $B_\perp = 15$~T and $B_\perp = 30$~T shown as purple and blue lines in Fig.~\ref{fig:out-of-plane-dts}\,(e) shifts the A$'$ and B$'$~excitons towards higher energies, cf. dotted lines in Fig.~\ref{fig:out-of-plane-dts}\,(a).
		Accordingly, the DTS also shift toward higher energies while the dispersive shapes remain qualitatively the same.
		Again, the A$-$A$'$ and A$-$B$'$ biexciton resonances {inherit} the $g$-factor of the A$'_\text{1s}$ and B$'_\text{1s}$~excitons, respectively.
		This is in contrast to a simple band analysis which suggests that the Zeeman shift of the A$^{\phantom{\prime}}_\text{1s}$ exciton is compensated by an opposite shift of the A$'_\text{1s}$ or B$'_\text{1s}$ exciton leading to an almost vanishing biexciton $g$-factor.
		While this compensation of $g$-factors applies to the A$-$A$'$ and A$-$B$'$ biexcitons $B\hspace{0.5mm}^{\text{A\smash[b]{,}A\textquotesingle}}_{-,\mu=b}$ and $B\hspace{0.5mm}^{\text{A\smash[b]{,}B\textquotesingle}}_{-,\mu=b}$, it does not hold true for the biexciton resonances $B\hspace{0.5mm}^{\text{A\smash[b]{,}A\textquotesingle}}_{-,\mu=b}(P\hspace{0.5mm}^{\text{A}}_{\text{1s}})^*$ and $B\hspace{0.5mm}^{\text{A\smash[b]{,}B\textquotesingle}}_{-,\mu=b}(P\hspace{0.5mm}^{\text{A}}_{\text{1s}})^*$.
		Therefore, the biexciton resonances $B\hspace{0.5mm}^{\text{A\smash[b]{,}A\textquotesingle}}_{-,\mu=b}(P\hspace{0.5mm}^{\text{A}}_{\text{1s}})^*$ and $B\hspace{0.5mm}^{\text{A\smash[b]{,}B\textquotesingle}}_{-,\mu=b}(P\hspace{0.5mm}^{\text{A}}_{\text{1s}})^*$ inherit the $g$-factors of the A$'_\text{1s}$ and B$'_\text{1s}$ exciton, respectively.
		In contrast, we expect a twice as large Zeeman shift of A$-$Ad$'$ and Ad$-$A$'$ biexciton resonances, cf. pale dash-dotted line Fig.~\ref{fig:out-of-plane-dts}\,(a).
		However, due to their negligible oscillator strengths in coherent pump-probe spectroscopy they appear not as resonances in Fig.~\ref{fig:out-of-plane-dts}\,(e). 
		Note that previous photoluminescence measurements \cite{ye2018efficient,barbone2018charge,li2018revealing} ascribed A$-$Ad$'$ and Ad$-$A$'$ biexciton resonances a smaller Zeeman shift which is similar to our expectations for the A$-$A$'$ biexciton resonance.
		{\textit{We have shown that the differential transmission spectra in the presence of an out-of-plane magnetic field $B_\perp$ mirror the Zeeman shifts.
		In particular, the $g$-factor of bright$-$bright biexciton resonances inherits the $g$-factor of the associated exciton resonance.}}
	
\subsection{In-plane Magnetic Field} \label{sub:in}

		\begin{figure*}
			\centering
			\includegraphics[width=1.95\columnwidth]{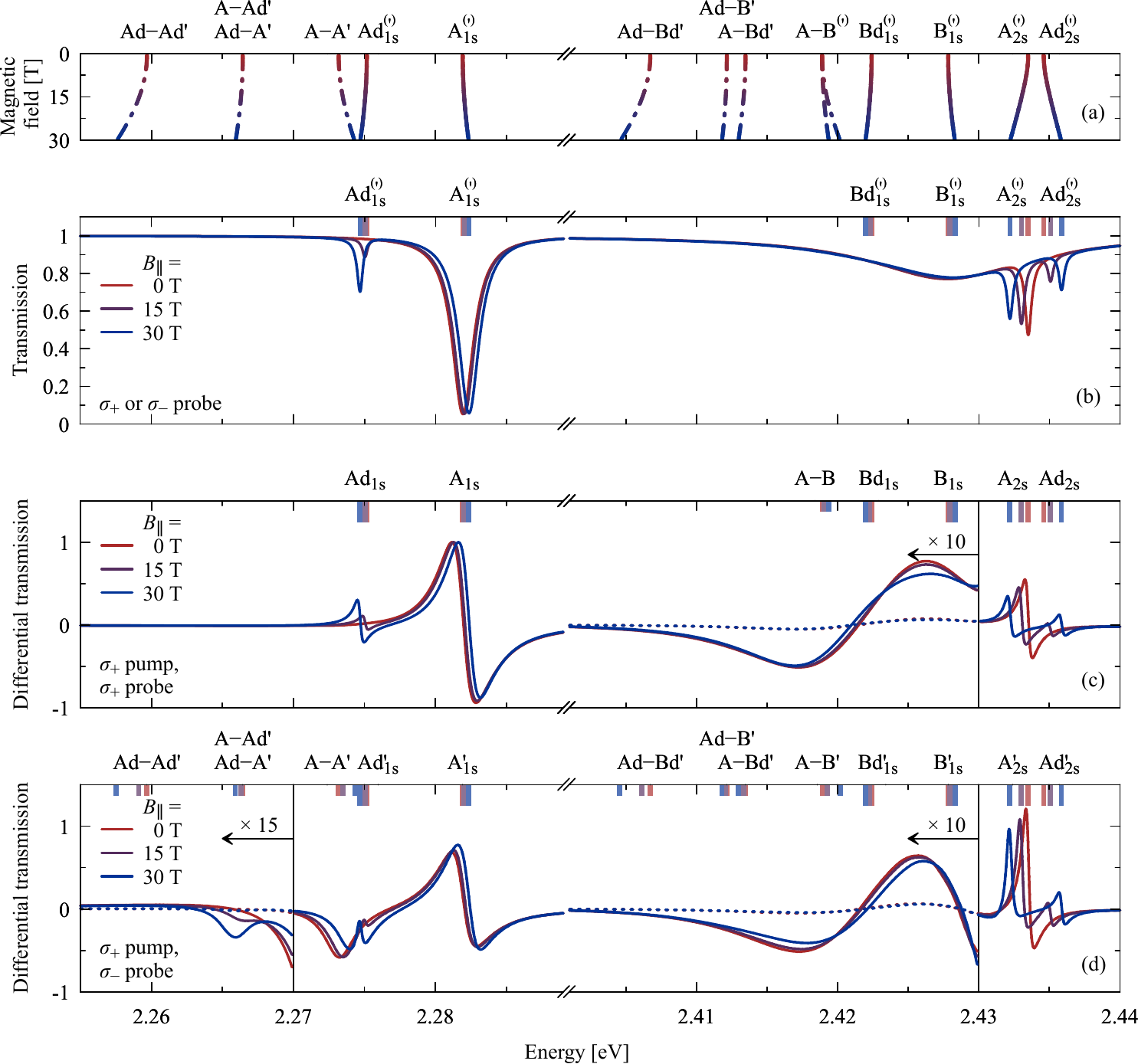}
			\caption{\textbf{In-plane magnetic-field-dependent transmission and differential transmission.}
				(a)~Exciton and biexciton resonance energies for different in-plane magnetic fields $B_\parallel$, cf. Figs.~\ref{Bild-AB} and \ref{fig:bix-schema}.
				Solid (dotted) lines represent exciton resonances and dashed (dash-dotted) lines biexciton resonances for $\sigma_{+}$ ($\sigma_-$) circularly polarized excitation.
				(b)~Linear transmission spectra of monolayer MoS$_2$ encapsulated in hBN at 5~K for different $B_\parallel$.
				(The transmission for $\sigma_+$ and $\sigma_-$~circularly polarized light is identical.)
				The $B_\parallel$ dependent exciton energies are indicated as colored bars above.
				(c,d)~Normalized differential transmission spectra for different magnetic fields $B_\parallel$.
				The A$_\text{1s}$~exciton is resonantly pumped by $\sigma_+$~circularly polarized light and the energetically broadband probe pulse is either (c)~$\sigma_+$~circularly polarized or (d)~$\sigma_-$~circularly polarized.
				The differential transmission is partially enhanced as indicated while the original signal is plotted by dotted lines.
				The colored long bars above mark the exciton energies, whereas the shorter bars indicate the biexciton energies.
			}
			\label{fig:in-plane-dts}
		\end{figure*}
		\textit{Linear response:}
		Next, we study the impact of an in-plane magnetic field $B_\parallel$.
		In contrast to an out-of-plane magnetic field $B_\perp$, an in-plane magnetic field $B_\parallel$ leads to identical shifts of A and A$'$ as well as B and B$'$~excitons in the linear regime drawn as solid lines in Fig.~\ref{fig:in-plane-dts}\,(a).
		This is due to a valley independent coupling among bright and dark excitons described by Eq.~(\ref{eq:in-plane-mixing}).
		Therefore, the linear response plotted in Fig.~\ref{fig:in-plane-dts}\,(b) is identical for $\sigma_+$ and $\sigma_-$~circularly polarized light.
		The linear transmission at $B_\parallel=0$~T is plotted as a red line in Fig.~\ref{fig:in-plane-dts}\,(b) and shows the A$_\text{1s}$, B$_\text{1s}$ and A$_\text{2s}$~exciton resonances.
		The $B_\parallel$ dependent bright-dark (spin-state) exciton mixing results in brightened dark excitons due to a redistribution of the oscillator strength between bright and dark excitons.
		The brightened dark Ad$_\text{1s}$ and Ad$_\text{2s}$~excitons appear as resonances in the transmission spectra at $B_\parallel=15$~T and $B_\parallel=30$~T plotted as purple and blue curves in Fig.~\ref{fig:in-plane-dts}\,(b), respectively.
		Note that the exciton state ordering of bright A$_{2s}$ and dark Ad$_{2s}$ excitons sensitively depends on the local-field intravalley exchange Coulomb interaction and might also be inverted for larger values of the exchange Coulomb potential given in Eq.~\eqref{eq:lf-effect}.
		The bright-dark level repulsion leads to energy renormalizations of the A$^{(\prime)}$, Ad$^{(\prime)}$~excitons with increasing $B_\parallel$ \cite{lu2019magnetic,robert2020measurement} shown by the solid lines in Fig.~\ref{fig:in-plane-dts}\,(a).
		The narrower dark exciton linewidths compared to bright excitons originate from suppressed radiative linewidth contributions.
		This is in agreement with experimental observations \cite{zhang2017magnetic,robert2017fine,zhou2017probing}.
		The dark exciton linewidths are dominated by the phonon-mediated linewidth contributions $2\gamma\hspace{0.5mm}^{\text{Ad}_\text{1s}}_{\text{x}} = 0.4$~meV at 5~K \cite{linewidth}.
		With increasing $B_\parallel$, the redistribution from bright to dark exciton oscillator strength increases the radiative Ad$_\text{1s}$~exciton linewidth contribution from $2\gamma\hspace{0.5mm}^{\text{Ad}_\text{1s}}_{\text{r}} = 0$~meV at $B_\parallel=0$~T to $2\gamma\hspace{0.5mm}^{\text{Ad}_\text{1s}}_{\text{r}} = 0.1$~meV at $B_\parallel= 30$~T.
		Thus, the total Ad$_\text{1s}$~exciton linewidth slightly rises from 0.4~meV to 0.5~meV.
		Simultaneously, the radiative A$_\text{1s}$~exciton linewidth slightly decreases from $2\gamma\hspace{0.5mm}^{\text{A}_\text{1s}}_{\text{r}} = 1.4$~meV at $B_\parallel= 0$~T to $2\gamma\hspace{0.5mm}^{\text{A}_\text{1s}}_{\text{r}} = 1.3$~meV at $B_\parallel= 30$~T and the total A$_\text{1s}$~linewidth declines from 1.8~meV to 1.7~meV.
		Even though the dark Bd$_\text{1s}$~exciton with a phonon-mediated linewidth contribution of $2\gamma\hspace{0.5mm}^{\text{Bd}_\text{1s}}_{\text{x}} = 11.4$~meV at 5~K is also excited, its appearance as a sharp resonance in Fig.~\ref{fig:in-plane-dts}\,(b) is obscured by the large B and Bd~exciton linewidths compared to their small energy separation.
		\textit{DTS for $\sigma_+$ pump and $\sigma_+$ probe pulses:}
		Compared to a zero in-plane magnetic field $B_\parallel = 0$~T, shown as red line in Fig.~\ref{fig:in-plane-dts}\,(c), an in-plane magnetic field of $B_\parallel = 15$~T or $B_\parallel=30$~T plotted as purple and blue lines in Fig.~\ref{fig:in-plane-dts}\,(c) leads to additional DTS features associated with dark Ad~excitons.
		These DTS features describe dispersive profiles which -- similar to the A$_\text{1s}$ and A$_\text{2s}$ response -- account for blue shifted dark Ad$_\text{1s}$ and Ad$_\text{2s}$ exciton resonances with exciton-scattering-induced sidebands, cf. Fig.~\ref{dts-schema}\,(a).
		The level repulsion between bright and dark excitons, cf. solid lines in Fig.~\ref{fig:in-plane-dts}\,(a), is also mirrored in the DTS shown by the purple and blue lines in Fig.~\ref{fig:in-plane-dts}\,(c).
		In contrast, the large B$_\text{1s}$ and Bd$_\text{1s}$~exciton linewidths compared to their small energy separation obscure the identification of DTS features from Bd$_\text{1s}$~excitons.
		\textit{DTS for $\sigma_+$ pump and $\sigma_-$ probe pulses:}
		With rising $B_\parallel$ from $B_\parallel=0$~T plotted as red line in Fig.~\ref{fig:in-plane-dts}\,(d) to $B_\parallel=15$~T and $B_\parallel=30$~T shown as purple and blue lines in Fig.~\ref{fig:in-plane-dts}\,(d), the pump-dependent renormalizations and redistributions of oscillator strengths of dark Ad$_\text{1s}'$~excitons yield dips superimposed on the negative DTS signal from A$-$A$'$ intervalley biexcitons (Fig.~\ref{fig:bix-schema}\,(b)).
		Similarly, the dispersive DTS profile at Ad$_\text{2s}'$ accounts for a blue shifted Ad$_\text{2s}'$~resonance with exciton-scattering-induced sidebands.
		The $B_\parallel$ dependent brightening of dark Ad$^{(\prime)}$ and Bd$^{(\prime)}$~excitons also results in additional intervalley biexciton resonances.
		In general, this includes Ad$-$Ad$'$, A$-$Ad$'$, Ad$-$A$'$, Bd$-$Bd$'$, B$-$Bd$'$, Bd$-$B$'$, Ad$-$Bd$'$, Ad$-$B$'$, A$-$Bd$'$, Bd$-$Ad$'$, B$-$Ad$'$, Bd$-$A$'$~intervalley biexcitons in addition to the A$-$A$'$, B$-$B$'$, A$-$B$'$, and B$-$A$'$~intervalley biexcitons, cf. Fig.~\ref{fig:bix-schema}.
		The intervalley biexciton fine structure which is relevant for resonantly pumping the A$_\text{1s}$~exciton is shown as dash-dotted lines in Fig.~\ref{fig:in-plane-dts}\,(a).
		The A$-$Ad$'$ and Ad$-$A$'$~intervalley biexcitons are visible as superimposed weak negative DTS signals below the A$-$A$'$ biexcitons.
		In contrast, A$-$Bd$'$ and Ad$-$B$'$~intervalley biexcitons are obscured due to their large phonon-mediated linewidths.
		On the other hand, Ad$-$Ad$'$ and Ad$-$Bd$'$~intervalley biexcitons exhibit very low oscillator strengths since they are only Coulomb driven by dark Ad, Ad$'$, and Bd$'$~excitons with much lower oscillator strengths than their bright counterparts.
		{\textit{Applying an in-plane magnetic field $B_\parallel$ brightens not only previously spin-forbidden dark exciton resonances but also a multitude of bright$-$dark and dark$-$dark biexciton resonances.
		Furthermore, an in-plane magnetic field $B_\parallel$ leads to a pronounced differential transmission signatures due to the renormalization of Ad$_\text{1s}$ and Ad$_\text{2s}$ excitons.}}

\subsection{Tilted Magnetic Field} \label{sub:tilted}

		\begin{figure*}
			\centering
			\includegraphics[width=1.95\columnwidth]{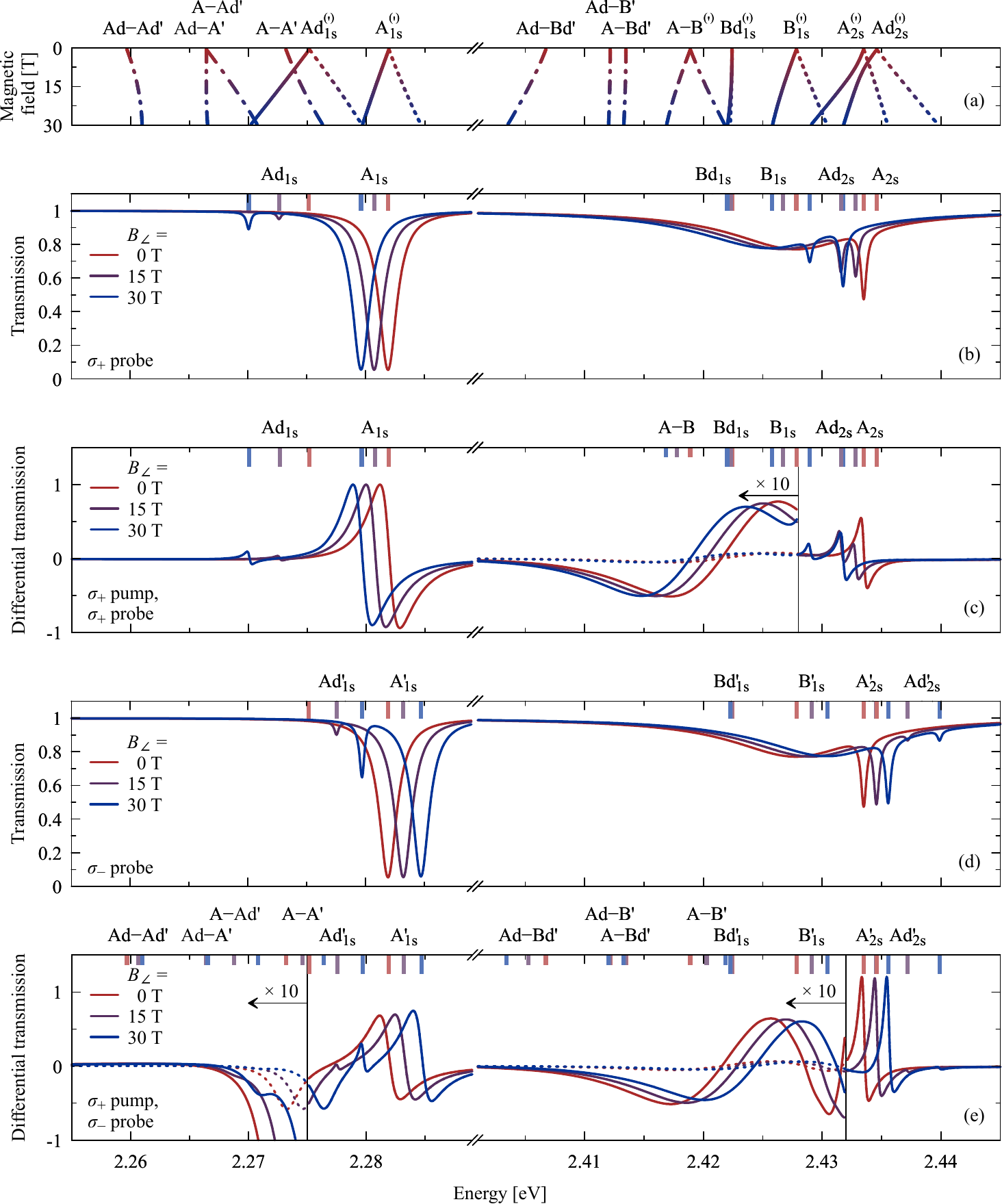}
			\caption{\textbf{Tilted magnetic-field-dependent transmission and differential transmission.}
				(a)~Exciton and biexciton resonance energies for different tilted magnetic fields $B_\angle$, cf. Figs.~\ref{Bild-AB} and \ref{fig:bix-schema}.
				Solid (dotted) lines represent exciton resonances and dashed (dash-dotted) lines biexciton resonances for $\sigma_{+}$ ($\sigma_-$) circularly polarized excitation.
				(b,d)~Linear transmission spectra of monolayer MoS$_2$ encapsulated in hBN at 5~K for (b)~$\sigma_+$ and (d)~$\sigma_-$~circularly polarized light for different $B_\angle$.
				The $B_\angle$ dependent exciton energies are indicated as colored bars above.
				(c,e)~Normalized differential transmission spectra for different magnetic fields $B_\angle$.
				The A$_\text{1s}$~exciton is resonantly pumped by $\sigma_+$~circularly polarized light and the energetically broadband probe pulse is either (c)~$\sigma_+$~circularly polarized or (e)~$\sigma_-$~circularly polarized.
				The differential transmission is partially enhanced as indicated while the original signal is plotted by dotted lines.
				The colored long bars above mark the exciton energies, whereas the shorter bars indicate the biexciton energies.
			}
			\label{fig:tilted-dts}
		\end{figure*}
		\textit{Linear response:}
		In the following, we study the influence of magnetic fields $B_\angle$ applied under a 45° tilt angle.
		This combines the previously discussed effects from out-of-plane $B_\perp$ and in-plane $B_\parallel$ fields.
		The $B_\angle$ dependent shifts of exciton resonances for monolayer MoS$_2$ are shown in Fig.~\ref{fig:tilted-dts}\,(a).
		Like the bright A$_\text{1s}$~transition, the dark Ad$_\text{1s}$~exciton shifts red but with approximately double magnitude increasing the bright-dark splitting $\Delta_{\text{A}_\text{1s}-\text{Ad}_\text{1s}}$ with rising $B_\angle>0$~T.
		The increasing $\Delta_{\text{A}_\text{1s}-\text{Ad}_\text{1s}}$ yields weakly pronounced Ad$_\text{1s}$~excitons in the $\sigma_+$~circularly polarized linear transmission at $B_\angle=15$~T and $B_\angle=30$~T plotted as blue and purple lines in Fig.~\ref{fig:tilted-dts}\,(b).
		Simultaneously, the bright-dark splitting $\Delta_{\text{A}_\text{1s}'-\text{Ad}_\text{1s}'}$ decreases for increasing $B_\angle>0$~T.
		This leads to strongly pronounced Ad$_\text{1s}'$~resonances in the $\sigma_-$~circularly polarized linear transmission at $B_\angle=15$~T and $B_\angle=30$~T shown as purple and blue curves in Fig.~\ref{fig:tilted-dts}\,(d).
		\textit{DTS for $\sigma_+$ pump and $\sigma_+$ probe pulses:}
		A tilted magnetic field $B_\angle$ combines the multitude of Zeeman shifts and brightened dark excitons.
		In particular, their combination allows to control the oscillator strengths of coherent signatures associated with dark excitons in pump-probe spectroscopy.
		Most notably, the increasing bright-dark splitting $\Delta_{\text{A}_\text{1s}-\text{Ad}_\text{1s}}$ from $B_\angle=0$~T to $B_\angle=15$~T to $B_\angle=30$~T induces weak Ad$_\text{1s}$~exciton DTS signals shown by the red to purple to blue plots Fig.~\ref{fig:tilted-dts}\,(c).
		Simultaneously, the DTS of the Ad$_\text{2s}$~exciton is strongly pronounced.
		In contrast, an oppositely oriented magnetic field $B_\angle<0$~T gives the opportunity to enhance the Ad$_\text{1s}$~exciton DTS response and decrease the DTS of the Ad$_\text{2s}$~exciton, cf. Appendix~\ref{sec:app-dts}.
		\textit{DTS for $\sigma_+$ pump and $\sigma_-$ probe pulses:}
		A tilted magnetic field of $B_\angle=15$~T and $B_\angle=30$~T, shown as purple and blue lines in Fig.~\ref{fig:tilted-dts}\,(e), leads to strong positive dips from Ad$_\text{1s}'$~excitons.
		The opposite holds true at $B_\angle<0$~T where DTS contributions from Ad$_\text{1s}'$~excitons are suppressed, cf. Appendix~\ref{sec:app-dts}.
		The large Ad$_\text{1s}'$~exciton oscillator strength also enhances the A$-$Ad$'$ biexciton oscillator strength compared to an in-plane field $B_\parallel$ solely as shown in Fig.~\ref{fig:in-plane-dts}\,(d).
		However, the increased A$-$Ad$'$ biexciton oscillator strength is accompanied by an increased background signal from the A$-$A$'$ biexciton.
		On the other hand, $B_\angle<0$~T results in an increased energy separation between A$-$Ad$'$ and A$-$A$'$ biexcitons and lower background signal from A$-$A$'$ biexcitons while the A$-$Ad$'$ biexciton oscillator strength decreases, cf. Appendix~\ref{sec:app-dts}.
		Furthermore, features from Ad$'_\text{2s}$~excitons are suppressed in Fig.~\ref{fig:tilted-dts}\,(e) due to their low oscillator strengths.
		The opposite holds true for negative tilted fields $B_\angle<0$~T, where the A$-$Ad$'$ biexcitons are more pronounced, cf. Appendix~\ref{sec:app-dts}.
		{\textit{We have shown that a tilted magnetic field $B_\angle$ allows to control the signal strength of the differential transmission associated with the renormalization of previously spin-forbidden dark excitons and bright$-$dark biexciton resonances.
		This originates from the combined influence of in-plane and out-of-plane magnetic field contributions which allows to enhance or suppress the corresponding pump-probe signal.}}

\section{Conclusion} \label{sec:conclusion}

		In conclusion, we have presented a microscopic description to access the coherent exciton kinetics in monolayer TMDCs in the presence of differently oriented magnetic fields.
		We provide the magnetic-field-dependent exciton and biexciton resonance energies, transmission spectra, and differential transmission spectra.
		In particular, the latter reveals the manipulation of exciton-exciton scattering by magnetic fields.
		Here, we focused on the scattering induced changes of the exciton resonances, calculated the biexciton oscillator strengths, and predicted the possibility to detect the corresponding biexcitons in optical wave-mixing spectroscopy.
		Thus, our results provide a roadmap to interpret coherent pump-probe spectra in the presence of external magnetic fields.

\begin{acknowledgments}

		All authors gratefully acknowledge funding from the Deutsche Forschungsgemeinschaft through Project No. 420760124 (KN 427/11-1 and BR 2888/8-1).
		F.K. thanks the Berlin School of Optical Sciences and Quantum Technology.

\end{acknowledgments}
\appendix
		
\section{Magnetic Field} \label{app:magnetic-field}

		\begin{figure*}
			\centering
			\includegraphics[width=2\columnwidth]{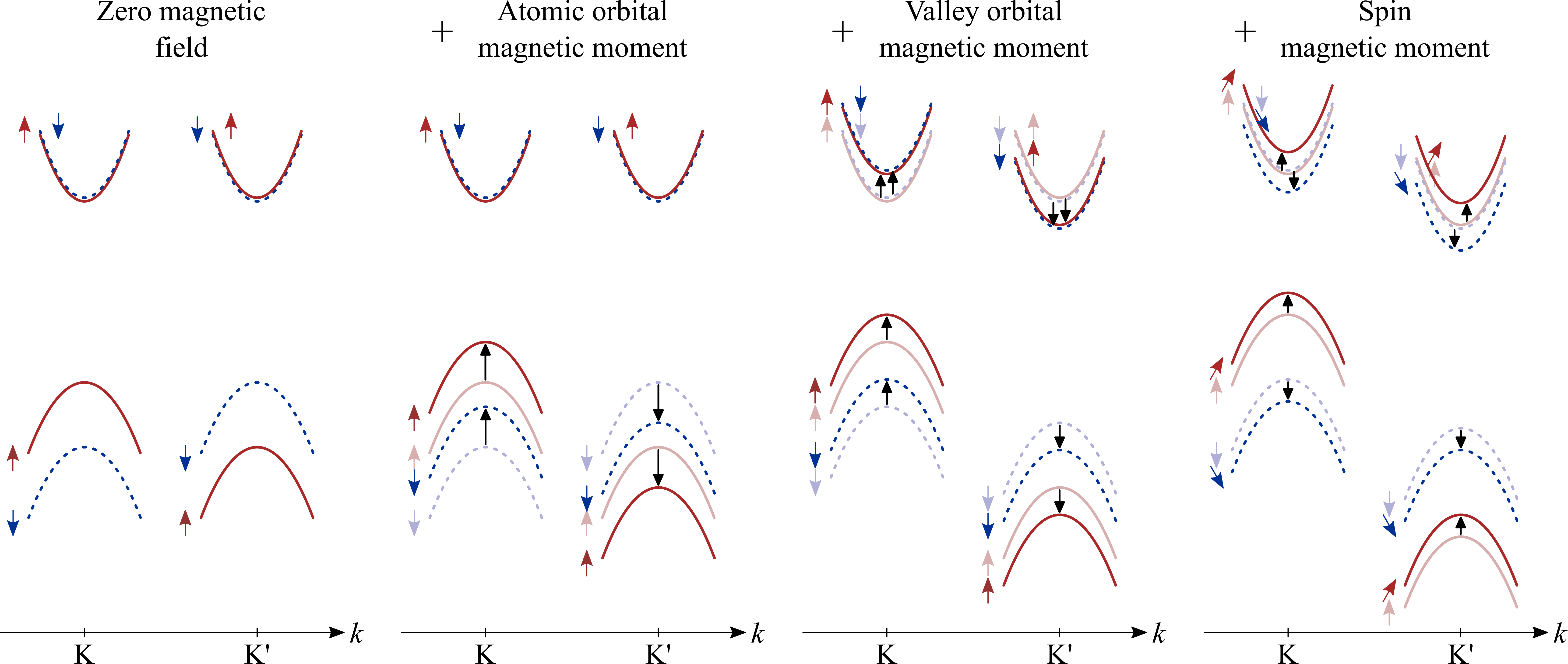}
			\caption{\textbf{Illustration of the contributing magnetic moments.}
				Shown is the single particle band structure for molebdenum based TMDCs.
				Red continuous lines characterize spin-$\uparrow$ bands and blue dashed lines are spin-$\downarrow$ bands.
				The influence of the different magnetic moments as illustrated by black arrows is successively added up.
				The atomic and valley orbital magnetic moments introduce Zeeman shifts.
				In addition to a shift, the spin magnetic moment results in a coupling of spin-allowed interband transitions with same electron and hole spin as well as spin-forbidden transitions with opposite electron and hole spin.
				This effectively resembles a softening of the spin selection rules as illustrated by tilted arrows.
			}
			\label{Bild-exp}
		\end{figure*}

		In this Appendix the different magnetic moments are individually discussed.
		Without loss of generality, the in-plane component of the magnetic field is assumed to be aligned along the x-direction $\textbf{\textit{B}} = (B_\parallel,0,B_\perp)^T$ with the in-plane $B_\parallel = B \sin \theta$ and out-of-plane contributions $B_\perp = B \cos \theta$.

		\subsection{Atomic Orbital Magnetic Moment} \label{sub:mag-1}

		The atomic orbital contribution of the Zeeman shift also referred to as intracellular orbital moment $m_l \mu_B B_\perp$ \cite{aivazian2015magnetic,srivastava2015valley} is determined by the magnetic quantum number~$m_l$, the Bohr magneton~$\mu_B$, and the magnetic field perpendicular to the monolayer sample~$B_\perp$.
		As the conduction bands are primarily constructed from hybridized $d_{z^2}$~orbitals with $l=m_l=0$ \cite{xiao2012coupled,cao2012valley,kuc2015electronic} the associated atomic orbital magnetic moment is negligible.
		In contrast, the valence bands arise from hybridization of $d_{x^2-y^2}$ and $d_{xy}$~orbitals $d_{x^2-y^2}+\text{i} d_{xy}$ with magnetic quantum number $m_l = 2$ for the K~valley and $d_{x^2-y^2}-\text{i} d_{xy}$ with $m_l=-2$ for the K$'$~valley \cite{zhu2011giant,xiao2012coupled,cao2012valley,kosmider2013large}.
		The resulting intracellular orbital magnetic moment is described by:
		\begin{align}
		H_{\text{atomic}} = \sum_{\xi,s,\textbf{\textit{q}}} 2 \left( \delta_{\xi,\text{K}} - \delta_{\xi,\text{K}'}\right)\mu_B B_\perp \ v^\dag_{\xi,s,\textbf{\textit{q}}} v^{\phantom{\dag}}_{\xi,s,\textbf{\textit{q}}} .
		\end{align}

		\subsection{Valley Orbital Magnetic Moment} \label{sub:mag-2}

		The Zeeman shift associated with the valley magnetic moment is determined by $\left( \delta_{\xi,\text{K}} - \delta_{\xi,\text{K}'}\right) \mu_Bm_0/(\bar{m}) B_\perp$ \cite{yao2008valley,xu2014spin}.
		$m_0$ denotes the free electron mass and $\bar{m} = \frac{1}{8} \sum_{\xi,s} (m^e_{\xi,s}+m^h_{\xi,s})$ is the mean effective mass of the eight band model.
		The valley orbital magnetic moment leads to shifts of all conduction and valence bands with identical magnitudes but in opposite directions for the K and K$'$~valley described by the Hamiltonian:
		\begin{align}
		H_{\text{valley}} = \ & \sum_{\lambda,\xi,s,\textbf{\textit{q}}} \left( \delta_{\xi,\text{K}} - \delta_{\xi,\text{K}'}\right) \frac{\mu_Bm_0}{\bar{m}} B_\perp \ \lambda^\dag_{\xi,s,\textbf{\textit{q}}} \lambda^{\phantom{\dag}}_{\xi,s,\textbf{\textit{q}}} \notag \\
		= \ & \sum_{\xi,s,\textbf{\textit{q}}} \left( \delta_{\xi,\text{K}} - \delta_{\xi,\text{K}'}\right) \frac{\mu_Bm_0}{\bar{m}} B_\perp \ \notag \\
		& \hspace{6.2mm} \times \left(c^\dag_{\xi,s,\textbf{\textit{q}}} c^{\phantom{\dag}}_{\xi,s,\textbf{\textit{q}}} + v^\dag_{\xi,s,\textbf{\textit{q}}} v^{\phantom{\dag}}_{\xi,s,\textbf{\textit{q}}}\right) .
		\end{align}

		\subsection{Spin Magnetic Moment} \label{sub:mag-3}

		The spin magnetic moment is determined by $1/2 g \mu_B \textbf{\textit{s}} \cdot \textbf{\textit{B}}$ where $\textbf{\textit{s}} = (s_x,s_y,s_z)^T$ denotes the vector of spin Pauli matrices $s_x$, $s_y$, and $s_z$.
		The spin magnetic moment leads to a contribution associated with the in-plane component of the magnetic field $1/2 g_e \mu_B s_x B_\parallel$ as well as a term connected with the out-of-plane magnetic field $1/2 g_e \mu_B s_z B_\perp$ \cite{gong2013magnetoelectric,van2018strong}.
		Here, the $g$-factor can be approximately described by the free electron $g$-factor $g_e \approx 2$ \cite{kormanyos2014spin} which is in excellent agreement with experimental measurements \cite{lu2019magnetic}.
		The associated Hamiltonian is given by:
		\begin{align}
		H_{\text{spin}} = \ & \sum_{\lambda,\xi,s,\textbf{\textit{q}}} \left(\delta_{s,\uparrow} - \delta_{s,\downarrow}\right) \mu_B B_\perp \ \lambda^\dag_{\xi,s,\textbf{\textit{q}}} \lambda^{\phantom{\dag}}_{\xi,s,\textbf{\textit{q}}} \notag \\
		& + \sum_{\lambda,\xi,s,\textbf{\textit{q}}} \mu_B B_\parallel \ \lambda^\dag_{\xi,s,\textbf{\textit{q}}} \lambda^{\phantom{\dag}}_{\xi,\bar{s},\textbf{\textit{q}}} , \label{eq:spin-magn}
		\end{align}
		with $\bar{s} \neq s$, i.e., $\bar{s} = \ \downarrow$ for $s = \ \uparrow$ and $\bar{s} = \ \uparrow$ for $s = \ \downarrow$.
		The first term on the right-hand side of Eq.~(\ref{eq:spin-magn}) describes a spin-dependent Zeeman shift of the conduction and valence bands in the presence of an out-of-plane magnetic field $B_\perp$.
		The second line of Eq.~(\ref{eq:spin-magn}) describes a spin-mixing of electrons in the presence of an in-plane magnetic field $B_\parallel$.

		\subsection{Total Hamiltonian} \label{sub:mag-4}

		The electronic Hamiltonian involving the atomic orbital, valley orbital, and spin magnetic moment is represented by:
		\begin{align}
		H_{B} = \ & H_{\text{atomic}} +H_{\text{valley}} +H_{\text{spin}} \notag \\
		= \
		& \sum_{\xi,s,\textbf{\textit{q}}} \varepsilon\hspace{0.5mm}^{\xi,s}_{c,B_\perp} \ c^\dag_{\xi,s,\textbf{\textit{q}}} c^{\phantom{\dag}}_{\xi,s,\textbf{\textit{q}}} + \sum_{\xi,s,\textbf{\textit{q}}} \varepsilon\hspace{0.5mm}^{\xi,s}_{v,B_\perp} \ v^\dag_{\xi,s,\textbf{\textit{q}}} v^{\phantom{\dag}}_{\xi,s,\textbf{\textit{q}}} \notag \\
		& + \sum_{\xi,\textbf{\textit{q}},s} \mu_B B_\parallel \ c^\dag_{\xi,s,\textbf{\textit{q}}} c^{\phantom{\dag}}_{\xi,\bar{s},\textbf{\textit{q}}} + \sum_{\xi,\textbf{\textit{q}},s} \mu_B B_\parallel \ v^\dag_{\xi,s,\textbf{\textit{q}}} v^{\phantom{\dag}}_{\xi,\bar{s},\textbf{\textit{q}}} ,
		\end{align}
		with the conduction $\varepsilon\hspace{0.5mm}^{\xi,s}_{c,B_\perp}$ and valence $\varepsilon\hspace{0.5mm}^{\xi,s}_{v,B_\perp}$ band Zeeman shifts defined in Eqs.~(\ref{eq:Zeeman-1}) and (\ref{eq:Zeeman-2}) which linearly increase with the out-of-plane magnetic field~$B_\perp$.
		The individual contributions due to the atomic orbital, valley orbital, and spin magnetic moment are illustrated in Fig.~\ref{Bild-exp}.
		
\section{Coulomb Matrices}
		
		The dipole matrix element $d\hspace{0.5mm}^{c,v}_{\xi,s}$ projected on normalized Jones vectors are defined by  \cite{xiao2012coupled}:
		\begin{align}
		d\hspace{0.5mm}^{c,v}_{\xi,s} = - \text{i} \frac{{\sqrt{2}e_0  a_0 t_0}}{\varepsilon^{\xi,s}_g} . \label{eq:dipole-mat}
		\end{align}
		Here, $a_0$ is the lattice constant, $t_0$ is the effective hopping integral, and $\varepsilon^{\xi,s}_g$ is the energy gap between conduction and valence bands at the $\xi$ point with spin $s$.

		The matrix element associated with direct Coulomb scattering on a Hartree--Fock level is given by:
		\begin{align}
		W\hspace{0.5mm}^{\xi_1,s_1,s_2,s_3,s_4}_{V,\nu_2,\nu_3,\nu_4,\nu_1}
		& = \sum_{\textbf{\textit{k}}_1,\textbf{\textit{k}}_2} V_{\textbf{\textit{k}}_1-\textbf{\textit{k}}_2} \ \varphi^*\hspace{0.5mm}^{\xi_1,s_3,s_2}_{\nu_2,\textbf{\textit{k}}_1} \ \varphi^*\hspace{0.5mm}^{\xi_1,s_1,s_4}_{\nu_3,\textbf{\textit{k}}_2} \notag \\
		& \hspace{10.9mm} \times \left(\varphi\hspace{0.5mm}^{\xi_1,s_3,s_4}_{\nu_4,\textbf{\textit{k}}_1} - \varphi\hspace{0.5mm}^{\xi_1,s_3,s_4}_{\nu_4,\textbf{\textit{k}}_2}\right) \notag \\
		& \hspace{10.9mm} \times \left(\varphi\hspace{0.5mm}^{\xi_1,s_1,s_2}_{\nu_1,\textbf{\textit{k}}_1} - \varphi\hspace{0.5mm}^{\xi_1,s_1,s_2}_{\nu_1,\textbf{\textit{k}}_2}\right) \label{eq:mat-el-dir} .
		\end{align}
		The screened Coulomb potential $V_{\textbf{\textit{k}}_1-\textbf{\textit{k}}_2}$ is defined by an analytical model \cite{trolle2017model} which agrees with results obtained from \textit{ab initio} calculations \cite{latini2015excitons,andersen2015dielectric,qiu2016screening,steinhoff2017exciton}.
		Exchange Coulomb interactions originating from a local field effect \cite{qiu2015nonanalyticity,guo2019exchange,deilmann2019finite} are described by \cite{katsch2020theory}:
		\begin{align}
		X^{\xi_1,s_1,s_2,\xi_1,s_3}_{\nu_2,\nu_1,\textbf{\textit{q}}_1,\textbf{\textit{q}}_2} = V \begin{array}{l} {\textit{\scriptsize c,v,c,v}} \vspace{-1.5mm} \\ {\textit{\scriptsize $\xi_1,\xi_1,\xi_1,\xi_1$}} \vspace{-1.5mm} \\ {\textit{\scriptsize $s_3,s_1,s_1,s_3$}} \vspace{-1.5mm} \\ {\textit{\scriptsize $\mathbf{0}$}} \end{array} \ \varphi^*\hspace{0.5mm}^{\xi_1,s_3,s_3}_{\nu_2,\textbf{\textit{q}}_2} \ \varphi\hspace{0.5mm}^{\xi_1,s_1,s_2}_{\nu_1,\textbf{\textit{q}}_1} \label{eq:mat-el-exc-lin} .
		\end{align}
		The electron-hole exchange Coulomb potential on the right-hand side of Eq.~(\ref{eq:mat-el-exc-lin}) is adjusted to density function theory which calculate the constant value~$C$ for $\{\text{K},\uparrow\}$~excitons  \cite{qiu2015nonanalyticity}.
		Therefore, we renormalize the value for $C$ by the $\{\text{K},\uparrow\}$~exciton corresponding wave function $\varphi\hspace{0.5mm}^{\text{K},\uparrow,\uparrow}_{\text{1s},\textbf{\textit{q}}}$:
		\begin{align}
		V \begin{array}{l} {\textit{\scriptsize c,v,c,v}} \vspace{-1.5mm} \\ {\textit{\scriptsize $\xi_1,\xi_1,\xi_1,\xi_1$}} \vspace{-1.5mm} \\ {\textit{\scriptsize $s_1,s_2,s_2,s_1$}} \vspace{-1.5mm} \\ {\textit{\scriptsize $\mathbf{0}$}} \end{array} \approx \frac{C_{\xi_1,s_1,s_2}}{\big|\sum_{\textbf{\textit{q}}} \varphi\hspace{0.5mm}^{\text{K},\uparrow,\uparrow}_{\text{1s},\textbf{\textit{q}}}\big|^2}.
		\label{eq:lf-effect}
		\end{align}
		The matrix element for linear and nonlinear exchange Coulomb scattering including the local-field exchange potential defined in Eq.~(\ref{eq:mat-el-exc-lin}) is given by:
		\begin{align}
		W\hspace{0.5mm}^{\xi_1,s_1,s_3}_{0,\nu_2,\nu_1} = \ & \sum_{\textbf{\textit{q}}_1,\textbf{\textit{q}}_2} X^{\xi_1,s_1,s_1,\xi_1,s_3}_{\nu_2,\nu_1,\textbf{\textit{q}}_1,\textbf{\textit{q}}_2} ,  \label{eq:mat-el-exc-0}  \\
		W\hspace{0.5mm}^{\xi_1,s_1,s_2,s_3,s_4}_{0,\nu_2,\nu_3,\nu_4,\nu_1} = \ & \sum_{\textbf{\textit{q}}_1,\textbf{\textit{q}}_2} X^{\xi_1,s_1,s_2,\xi_1,s_4}_{\nu_2,\nu_1,\textbf{\textit{q}}_1,\textbf{\textit{q}}_2} \ \varphi^*\hspace{0.5mm}^{\xi_1,s_3,s_2}_{\nu_3,\textbf{\textit{q}}_1} \ \varphi\hspace{0.5mm}^{\xi_1,s_3,s_1}_{\nu_4,\textbf{\textit{q}}_1} \label{eq:mat-el-exc-1} .
		\end{align}
		Nonlinear exchange Coulomb interactions due to a nonlocal-field effect are described by:
		\begin{align}
		W\hspace{0.5mm}^{\xi_1,s_1,s_2,s_3,s_4}_{X,\nu_2,\nu_3,\nu_4,\nu_1} = \ & \sum_{\textbf{\textit{q}}_1,\textbf{\textit{q}}_2} V^{\xi_1,s_3,\xi_1,s_1}_{\text{X},\textbf{\textit{q}}_1-\textbf{\textit{q}}_2} \ \varphi^*\hspace{0.5mm}^{\xi_1,s_3,s_2}_{\nu_2,\textbf{\textit{q}}_1} \ \varphi^*\hspace{0.5mm}^{\xi_1,s_4,s_3}_{\nu_3,\textbf{\textit{q}}_2} \notag \\
		& \hspace{6.8mm} \times \varphi\hspace{0.5mm}^{\xi_1,s_4,s_1}_{\nu_4,\textbf{\textit{q}}_2} \ \varphi\hspace{0.5mm}^{\xi_1,s_1,s_2}_{\nu_1,\textbf{\textit{q}}_1} \label{eq:mat-el-exc-2} .
		\end{align}
		The involved exchange Coulomb matrix element $V^{\xi_1,s_3,\xi_1,s_1}_{\text{X},\textbf{\textit{q}}_1-\textbf{\textit{q}}_2}$ is defined according to:
		\begin{align}
		V^{\xi_1,s_1,\xi_2,s_2}_{\text{X},\textbf{\textit{q}}} \approx \ 
		& \delta_{\xi_1,\xi_2} V \begin{array}{l} {\textit{\scriptsize c,v,c,v}} \vspace{-1.5mm} \\ {\textit{\scriptsize $\xi_1,\xi_1,\xi_1,\xi_1$}} \vspace{-1.5mm} \\ {\textit{\scriptsize $s_1,s_2,s_2,s_1$}} \vspace{-1.5mm} \\ {\textit{\scriptsize $\mathbf{0}$}} \end{array} \notag \\
		&+
		\frac{V \left(\textbf{\textit{q}}\right)}{\varepsilon\left(\textbf{\textit{q}}\right)} \frac{1}{2 e_0^2} \ d\hspace{0.5mm}^{c,v}_{\xi_1,s_1} \left(d\hspace{0.5mm}^{c,v}_{\xi_2,s_2}\right)^* \, q_{\xi_1}^* q_{\xi_2}^{\vphantom{*}}
		\label{eq:low-wave-numb},
		\end{align}
		where $e_0>0$ denotes the elementary charge, the two wave vectors $q_{\xi_1}^*$, $q_{\xi_2}^{\vphantom{*}}$ are defined by $q_{\xi} = q_x + \text{i} \left(\delta_{\xi,\text{K}}-\delta_{\xi,\text{K\textquotesingle}}\right)  q_y $ \cite{kormanyos2015k}, and the dipole matrix elements $d\hspace{0.5mm}^{c,v}_{\xi_1,s_1}$, $d\hspace{0.5mm}^{c,v}_{\xi_2,s_2}$ were defined in Eq.~(\ref{eq:dipole-mat}).

		The two-electron and two-hole Coulomb interaction kernel for the two-electron and two-hole Schrödinger equation is defined by:
		\begin{align}
		& \hat{W}\hspace{0.5mm}^{\xi_1,s_1,\xi_2,s_2}_{\pm,\nu_1,\nu_2,\nu_3,\nu_4,\textbf{\textit{P}},\textbf{\textit{K}}} \notag \\
		& = V_{\textbf{\textit{P}}-\textbf{\textit{K}}} \sum_{\textbf{\textit{k}}_1} \varphi\hspace{0.5mm}^{\xi_1,s_1,s_1}_{\nu_1,\textbf{\textit{k}}_1} \Big(\varphi^*\hspace{0.5mm}^{\xi_1,s_1,s_1}_{\nu_3,\textbf{\textit{k}}_1-\beta_{\xi_1,s_1}(\textbf{\textit{P}}-\textbf{\textit{K}})} \notag \\
		& \hspace{34mm} - \varphi^*\hspace{0.5mm}^{\xi_1,s_1,s_1}_{\nu_3,\textbf{\textit{k}}_1+\alpha_{\xi_1,s_1}(\textbf{\textit{P}}-\textbf{\textit{K}})}\Big) \notag \\
		& \hspace{10.1mm} \times \sum_{\textbf{\textit{k}}_2} \varphi\hspace{0.5mm}^{\xi_2,s_2,s_2}_{\nu_2,\textbf{\textit{k}}_2} \Big(\varphi^*\hspace{0.5mm}^{\xi_2,s_2,s_2}_{\nu_4,\textbf{\textit{k}}_2+\beta_{\xi_2,s_2}(\textbf{\textit{P}}-\textbf{\textit{K}})} \notag \\
		& \hspace{34mm}- \varphi^*\hspace{0.5mm}^{\xi_2,s_2,s_2}_{\nu_4,\textbf{\textit{k}}_2-\alpha_{\xi_2,s_2}(\textbf{\textit{P}}-\textbf{\textit{K}})}\Big) \notag \\
		& \hspace{3.8mm} \pm \sum_{\textbf{\textit{k}}_1,\textbf{\textit{k}}_2} V_{\textbf{\textit{k}}_1-\textbf{\textit{k}}_2+\left(\alpha_{\xi_1,s_1}-\beta_{\xi_2,s_2}\right)\textbf{\textit{P}}+\textbf{\textit{K}}} \ \varphi\hspace{0.5mm}^{\xi_1,s_1,s_1}_{\nu_1,\textbf{\textit{k}}_1} \ \varphi\hspace{0.5mm}^{\xi_2,s_2,s_2}_{\nu_2,\textbf{\textit{k}}_2} \notag \\
		& \hspace{14.4mm} \times \Big(\varphi^*\hspace{0.5mm}^{\xi_1,s_1,s_1}_{\nu_3,\textbf{\textit{k}}_1-\beta_{\xi_1,s_1}(\textbf{\textit{P}}-\textbf{\textit{K}})} \notag \\
		& \hspace{19.8mm}- \varphi^*\hspace{0.5mm}^{\xi_1,s_1,s_1}_{\nu_3,\textbf{\textit{k}}_2-\alpha_{\xi_2,s_2}\textbf{\textit{P}}-\alpha_{\xi_1,s_1}\textbf{\textit{K}}}\Big) \notag \\
		& \hspace{14.4mm} \times \Big(\varphi^*\hspace{0.5mm}^{\xi_2,s_2,s_2}_{\nu_4,\textbf{\textit{k}}_1+\alpha_{\xi_1,s_1}\textbf{\textit{P}}+\alpha_{\xi_2,s_2}\textbf{\textit{K}}} \notag \\
		& \hspace{19.8mm} - \varphi^*\hspace{0.5mm}^{\xi_2,s_2,s_2}_{\nu_4,\textbf{\textit{k}}_2+\beta_{\xi_2,s_2}(\textbf{\textit{P}}-\textbf{\textit{K}})} \Big)  \label{eq:mat-el-bix-dir} .
		\end{align}
		The exchange Coulomb exciton-exciton scattering matrix is given by:
		\begin{align}
		& \hat{X}\hspace{0.5mm}^{\xi_1,s_5,s_2,\xi_2,s_3}_{\pm,\nu_3,\nu_4,\nu_5,\nu_6,\textbf{\textit{P}}} \notag \\
		& = V^{\xi_1,s_5,\xi_2,s_3}_{\text{X},\textbf{\textit{P}}} \sum_{\textbf{\textit{k}}_1} \varphi\hspace{0.5mm}^{\xi_1,s_2,s_2}_{\nu_3,\textbf{\textit{k}}_1} \ \varphi^*\hspace{0.5mm}^{\xi_1,s_2,s_2}_{\nu_5,\textbf{\textit{k}}_1-\beta_{\xi_1,s_2}\textbf{\textit{P}}} \notag \\
		& \hspace{26mm} \times \varphi^*\hspace{0.5mm}^{\xi_1,s_5,s_5}_{\nu_6,\textbf{\textit{k}}_1+\alpha_{\xi_1,s_2}\textbf{\textit{P}}} \sum_{\textbf{\textit{k}}_2} \varphi\hspace{0.5mm}^{\xi_2,s_3,s_3}_{\nu_4,\textbf{\textit{k}}_2} \notag \\
		& \hspace{4mm}\mp \sum_{\textbf{\textit{k}}_1,\textbf{\textit{k}}_2} V^{\xi_1,s_5,\xi_2,s_3}_{\text{X},\textbf{\textit{k}}_1-\textbf{\textit{k}}_2+\left(\alpha_{\xi_1,s_2}-\beta_{\xi_2,s_3}\right)\textbf{\textit{P}}} \ \varphi\hspace{0.5mm}^{\xi_1,s_2,s_2}_{\nu_3,\textbf{\textit{k}}_1} \  \varphi\hspace{0.5mm}^{\xi_2,s_3,s_3}_{\nu_4,\textbf{\textit{k}}_2} \notag \\
		&\hspace{14.6mm} \times \varphi^*\hspace{0.5mm}^{\xi_1,s_2,s_2}_{\nu_5,\textbf{\textit{k}}_1-\beta_{\xi_1,s_2}\textbf{\textit{P}}} \ \varphi^*\hspace{0.5mm}^{\xi_1,s_5,s_5}_{\nu_6,\textbf{\textit{k}}_2-\alpha_{\xi_2,s_3}\textbf{\textit{P}}} . \label{eq:mat-el-bix-exc}
		\end{align}
		Finally, the overlap matrix which directly appears due to the definition of the two-electron and two-hole correlation function in Eq.~(\ref{eq:four-particle-corr}) reads:
		\begin{align}
		& S\hspace{0.5mm}^{\xi_1,s_2,\xi_2,s_4}_{\pm,\nu_1,\nu_2,\nu_3,\nu_4,\textbf{\textit{P}},\textbf{\textit{K}}} \notag \\
		& = \delta_{\textbf{\textit{P}},\textbf{\textit{K}}} \mp \sum_{\textbf{\textit{k}}} \varphi\hspace{0.5mm}^{\xi_1,s_2,s_2}_{\nu_1,\textbf{\textit{k}}} \ \varphi\hspace{0.5mm}^{\xi_2,s_4,s_4}_{\nu_2,\textbf{\textit{k}}+\left(\alpha_{\xi_1,s_2}-\beta_{\xi_2,s_4}\right)\textbf{\textit{P}}+\textbf{\textit{K}}} \notag \\
		& \hspace{20.9mm} \times \varphi^*\hspace{0.5mm}^{\xi_1,s_2,s_2}_{\nu_3,\textbf{\textit{k}} -\beta_{\xi_1,s_2}\left(\textbf{\textit{P}}-\textbf{\textit{K}}\right)} \notag \\
		& \hspace{20.9mm} \times \varphi^*\hspace{0.5mm}^{\xi_2,s_4}_{\nu_4,\textbf{\textit{k}} +\alpha_{\xi_1,s_2}\textbf{\textit{P}} +\alpha_{\xi_2,s_4}\textbf{\textit{K}}} . \label{eq:S-Matrix}
		\end{align}

		\begin{figure*}
			\centering
			\includegraphics[width=1.95\columnwidth]{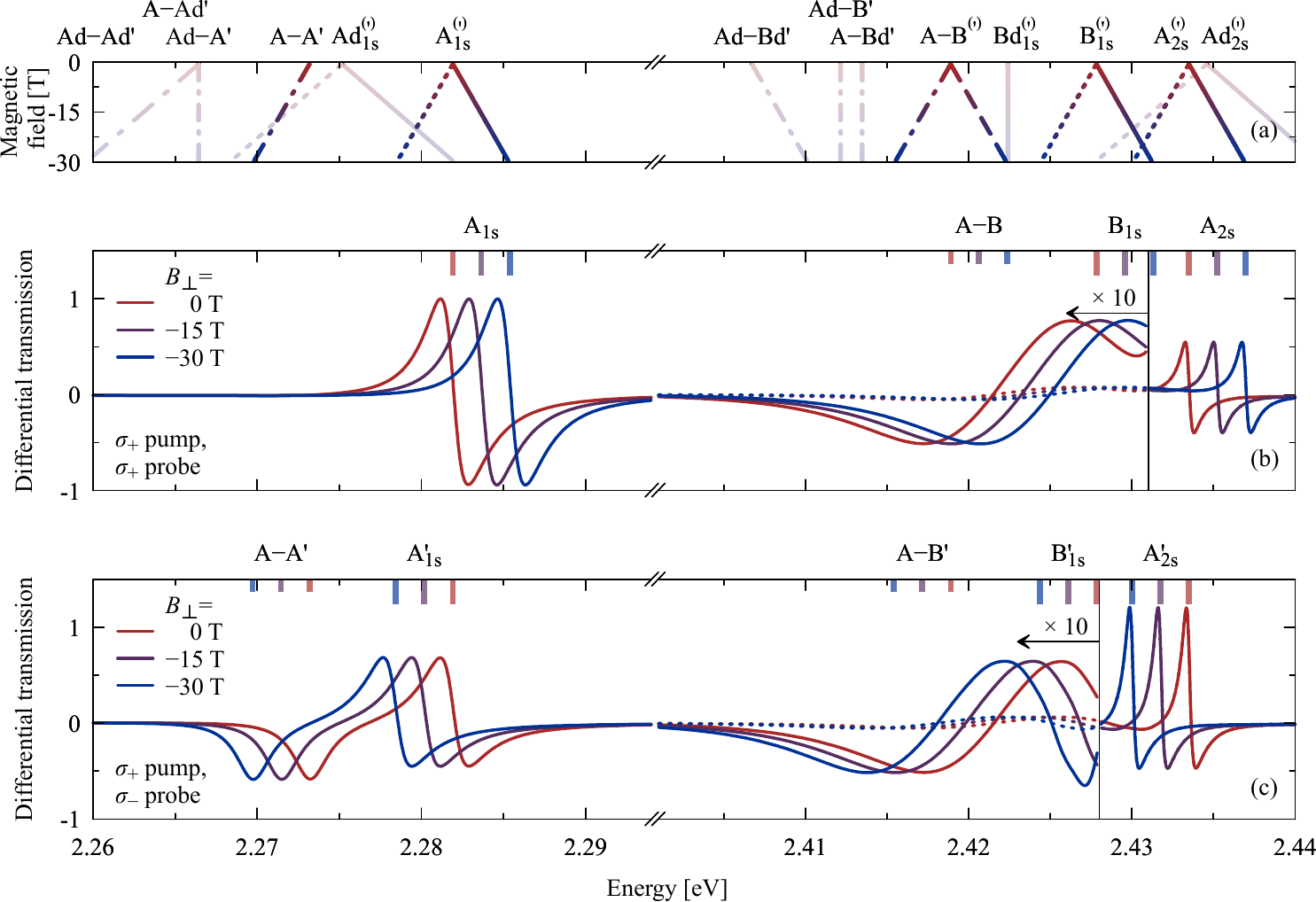}
			\caption{\textbf{Out-of-plane magnetic-field-dependent differential transmission.}
				Same as in Fig.~\ref{fig:out-of-plane-dts} but for negative~magnetic fields $B_\perp$ perpendicular to the monolayer sample.
			}
			\label{fig:dts-out-of-plane-dts-app}
		\end{figure*}
		\begin{figure*}
			\centering
			\includegraphics[width=1.95\columnwidth]{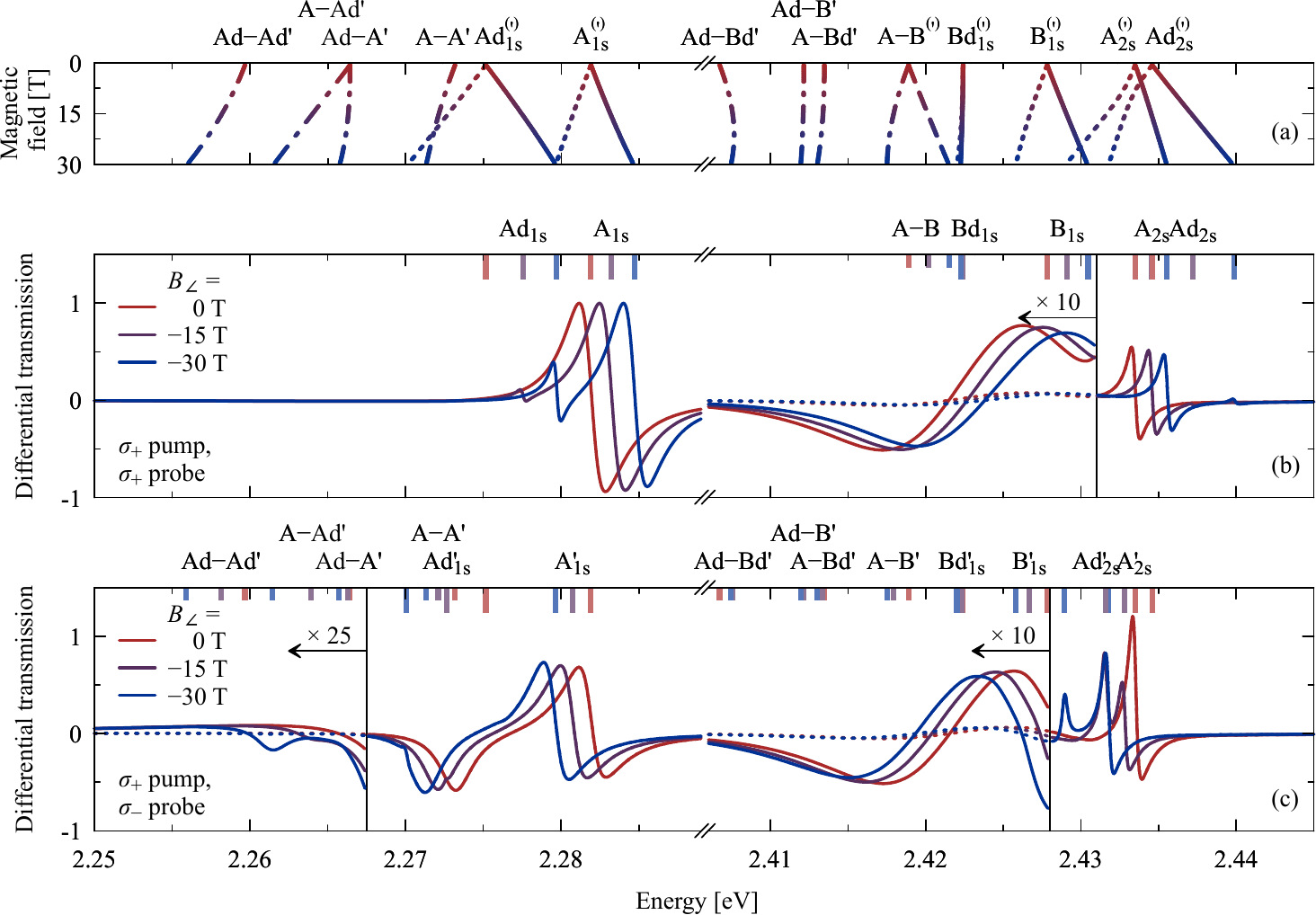}
			\caption{\textbf{Tilted magnetic-field-dependent differential transmission.}
				Same as in Fig.~\ref{fig:tilted-dts} but for negative~tilted magnetic fields $B_\angle$.
			}
			\label{fig:tilted-dts-app}
		\end{figure*}	
\section{Negative Magnetic Fields} \label{sec:app-dts}

		In the following, the results for negative magnetic fields are discussed.
		These results are equivalent to positive magnetic fields for pumping with $\sigma_-$~circularly polarized light and probing with same $\sigma_-$ or oppositely $\sigma_+$~circularly polarized light.
		The DTS for negative out-of-plane fields $B_\perp<0$~T is shown in Fig.~\ref{fig:dts-out-of-plane-dts-app}.
		In contrast to the response for positive magnetic fields $B_\perp>0$~T plotted in Fig.~\ref{fig:out-of-plane-dts}, the DTS signal in Fig.~\ref{fig:dts-out-of-plane-dts-app} mirrors the Zeeman shifts with opposite sign.
		Apart from this, the results closely resemble the positive magnetic fields and a corresponding interpretation holds true.
		As the DTS for an in-plane magnetic field with opposite sign is identical to the results in Fig.~\ref{fig:in-plane-dts}, we directly move on to discuss the case of a titled magnetic field $B_\angle<0$~T as shown in Fig.~\ref{fig:tilted-dts-app}.
		Compared to Fig.~\ref{fig:tilted-dts}\,(c), more pronounced Ad$_\text{1s}$ and less pronounced Ad$_\text{2s}$ features appear in Fig.~\ref{fig:tilted-dts-app}\,(b) for pumping and probing with $\sigma_+$~circularly polarized light.
		On the other hand, for a $\sigma_+$~circularly polarized pump pulse and a $\sigma_-$~circularly polarized probe pulse, the A$-$A$'$~intervalley biexciton and Ad$_\text{1s}'$~resonance show an energy crossing which gives an interfering signal, cf. Fig.~\ref{fig:tilted-dts-app}\,(c).
		Additionally, Fig.~\ref{fig:tilted-dts-app}\,(c) shows less pronounced DTS from A$-$Ad$'$ biexcitons compared to Fig.~\ref{fig:tilted-dts}\,(e).
		%
		

		%

\end{document}